\documentclass[fleqn,usenatbib]{mnras}
\usepackage{newtxtext,newtxmath}
 
\usepackage[T1]{fontenc}

\DeclareRobustCommand{\VAN}[3]{#2}
\let\VANthebibliography\thebibliography
\def\thebibliography{\DeclareRobustCommand{\VAN}[3]{##3}\VANthebibliography}

\usepackage{graphicx}	
\usepackage{amsmath}	
\usepackage{comment}
\usepackage{xcolor}
\usepackage[normalem]{ulem} 
\usepackage[flushleft]{threeparttable}
\usepackage{float}

\title[M87 jet observed by the VLBA at 8\&15\,GHz]{Properties of the jet in M87 revealed by its helical structure imaged with the
VLBA at 8 and 15\,GHz}

\author[Nikonov et al.]{\parbox{\textwidth}{
A.~S. Nikonov$^{1,2}$\thanks{E-mail: anikonov@mpifr-bonn.mpg.de \\ Member of the International Max Planck Research School (IMPRS) for Astronomy and Astrophysics at the Universities of Bonn and Cologne.},
Y.~Y. Kovalev$^{1,2,3}$,
E.~V. Kravchenko$^{3,2}$,
I.~N. Pashchenko$^{2}$,
A.~P. Lobanov$^{1,3}$
}
\vspace{0.4cm}\\
\parbox{\textwidth}{
$^{1}$Max-Planck-Institut f\"ur Radioastronomie, Auf dem H\"ugel 69, 53121, Bonn, Germany
\\
$^{2}$Lebedev Physical Institute of the Russian Academy of Sciences, Leninsky prospekt 53, 119991 Moscow, Russia
\\
$^{3}$Moscow Institute of Physics and Technology, Institutsky per.~9, Moscow region, Dolgoprudny, 141700, Russia
}}

\date{Accepted 2023 October 03. Received 2023 September 27; in original form 2023 March 25}

\pubyear{2023}

\begin{document}
\label{firstpage}
\pagerange{\pageref{firstpage}--\pageref{lastpage}}
\maketitle

\begin{abstract}
We present full-track high-resolution radio observations of the jet of the galaxy M87 at 8 and 15\,GHz. These observations were taken over three consecutive days in May 2009 using the Very Long Baseline Array (VLBA), one antenna of the Very Large Array (VLA), and the Effelsberg 100~m telescope. Our produced images have dynamic ranges exceeding 20,000:1 and resolve linear scales down to approximately 100 Schwarzschild radii, revealing a limb-brightened jet and a faint, steep spectrum counter-jet. We performed jet-to-counter-jet analysis, which helped estimate the physical parameters of the flow. The rich internal structure of the jet is dominated by three helical threads, likely produced by the Kelvin-Helmholtz (KH) instability developing in a supersonic flow with a Mach number of approximately 20 and an enthalpy ratio of around 0.3. We produce a $\texttt{CLEAN}$ imaging bias-corrected 8--15\,GHz spectral index image, which shows spectrum flattening in regions of helical thread intersections. This further supports the KH origin of the observed internal structure of the jet. We detect polarised emission in the jet at distances of approximately 20 milliarcseconds from the core and find Faraday rotation which follows a transverse gradient across the jet. We apply Faraday rotation correction to the polarisation position angle and find that the position angle changes as a function of distance from the jet axis, which suggests the presence of a helical magnetic field. 
\end{abstract}

\begin{keywords}
galaxies: active -- 
galaxies: jets -- 
techniques: interferometric -- 
galaxies: individual: M87
\end{keywords}


\section{Introduction}

M87 (Virgo~A, NGC~4486, 3C~274) is a giant elliptical galaxy which has an Active Galactic Nucleus (AGN) and an extended relativistic jet powered by a supermassive black hole (SMBH). The combination of proximity and SMBH makes M87 a prime target to study the outflow nature. The distance and the redsift for M87 are $D=$16.7\,Mpc \citep{Mei2007} and $z = 0.00436$ \citep{2000MNRAS.313..469S} correspondingly, which correspond to the angular scale of 1~mas $\approx0.08$\,pc. The supermassive black hole mass obtained from the Event Horizon Telescope image analysis is $(6.5 \pm 0.2|_\mathrm{stat} \pm 0.7|_\mathrm{sys}) \times 10^9 M_{\sun}$ \citep{EHT2019}. Thus, the Schwarzschild radius $R_\mathrm{s} \equiv 2GM/c^2 \approx 1.9 \times 10^{15}$~cm, which corresponds to 1~mas $\approx130R_\mathrm{s}$. 

The relativistic jet in M87 is frequently observed with the Very Long Baseline Interferometry (VLBI) technique. The images typically show an edge-brightened structure and a faint feature on the counter-jet side, south-east of the core. The full track 15~GHz observations which were done using the Very Large Baseline Array (VLBA) clearly show this feature \citep[e.g.][]{2007ApJ...660..200L,Kovalev2007}; see also the collection of MOJAVE images \citep{2018ApJS..234...12L}. This structure is believed to be the counter-jet deboosted by relativistic aberration effects. But there is no spectral information to support this idea. Spectral index maps could also help to better understand the nature of the edge-brightened \citep[e.g.][]{2008ASPC..386..155K} and triple-ridge jet structure \citep[e.g.][]{2016ApJ...833...56A, 2017Galax...5....2H,2021cosp...43E1398S}. The high-fidelity Very Large Array (VLA) radio images of the kiloparsec (kpc) scale jet of M87 clearly show a helical pattern \citep{Pasetto_2021}. This structure can also be seen in other jets, like 3C\,273 \citep{doi:10.1126/science.1063239} or S5~0836+710 \citep{2019A&A...627A..79V}. The oscillatory pattern of the kpc-scale jet of M87 was studied in \citet{2003NewAR..47..629L}, where the observed threads were identified with the Kelvin-Helmholtz (KH) instability happening in the jet. To our knowledge, no helical pattern is found and analysed in the inner, parsec scale jet, and high dynamic range images are needed to study plasma instability. 

\cite{Kovalev2007} also showed an extended high linear polarisation region but the absence of Faraday rotation maps did not allow the authors to reconstruct a magnetic field direction. Deep observations can significantly complement the polarisation studies of the M87 jet and reveal new regions not visible on \citet{2002ApJ...566L...9Z}, \citet{Park2019} and \citet{10.1093/mnras/stad525} polarisation maps.

These questions can be addressed using multi-frequency high-resolution observations. In this paper, we report on dual-frequency full-track high-sensitivity VLBA observations toward the M87 jet which have happened over three consecutive days. 
The paper is structured as follows.
\autoref{sec:obs} describes observations, data calibration and imaging processes. 
The analysis of the M87 jet structure is given in \autoref{sec:jet_structure}.
The spectral index analysis is presented in \autoref{sec:sp_ind}. In \autoref{sec:RM}, we present the linear polarisation results. The KH instability and the intrinsic jet parameters analyses are given in \autoref{sec:plasma_instability} and \autoref{sec:jet_to_counter-jet_flux_ratio}. 

\section{Observations and imaging}
\label{sec:obs}

\begin{figure*}
    \centering
    \includegraphics[width=\linewidth, trim=2cm 0cm 1.3cm 1.6cm, clip]{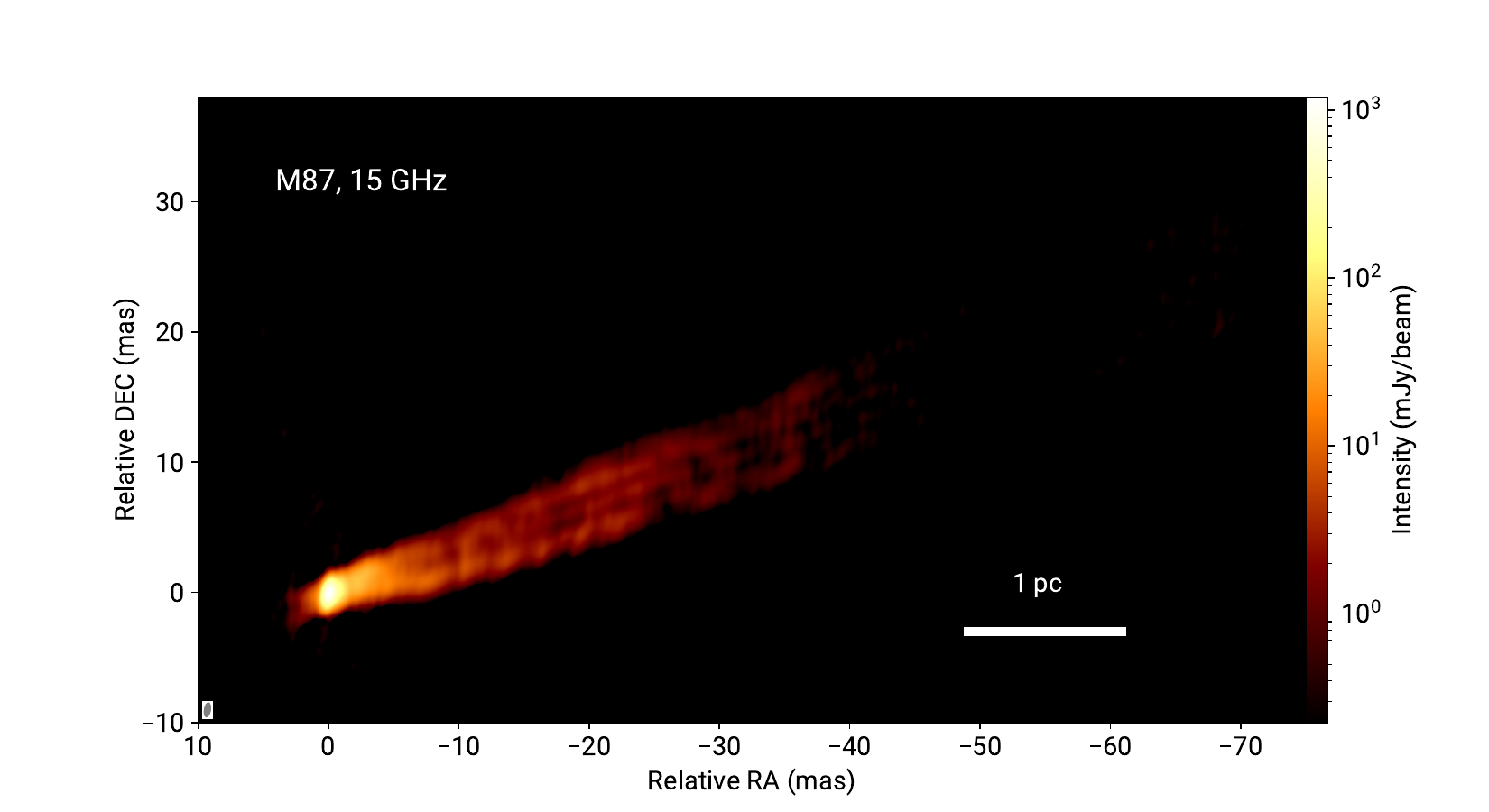}
    \includegraphics[width=\linewidth, trim=2cm 0.2cm 1.3cm 1.6cm, clip]{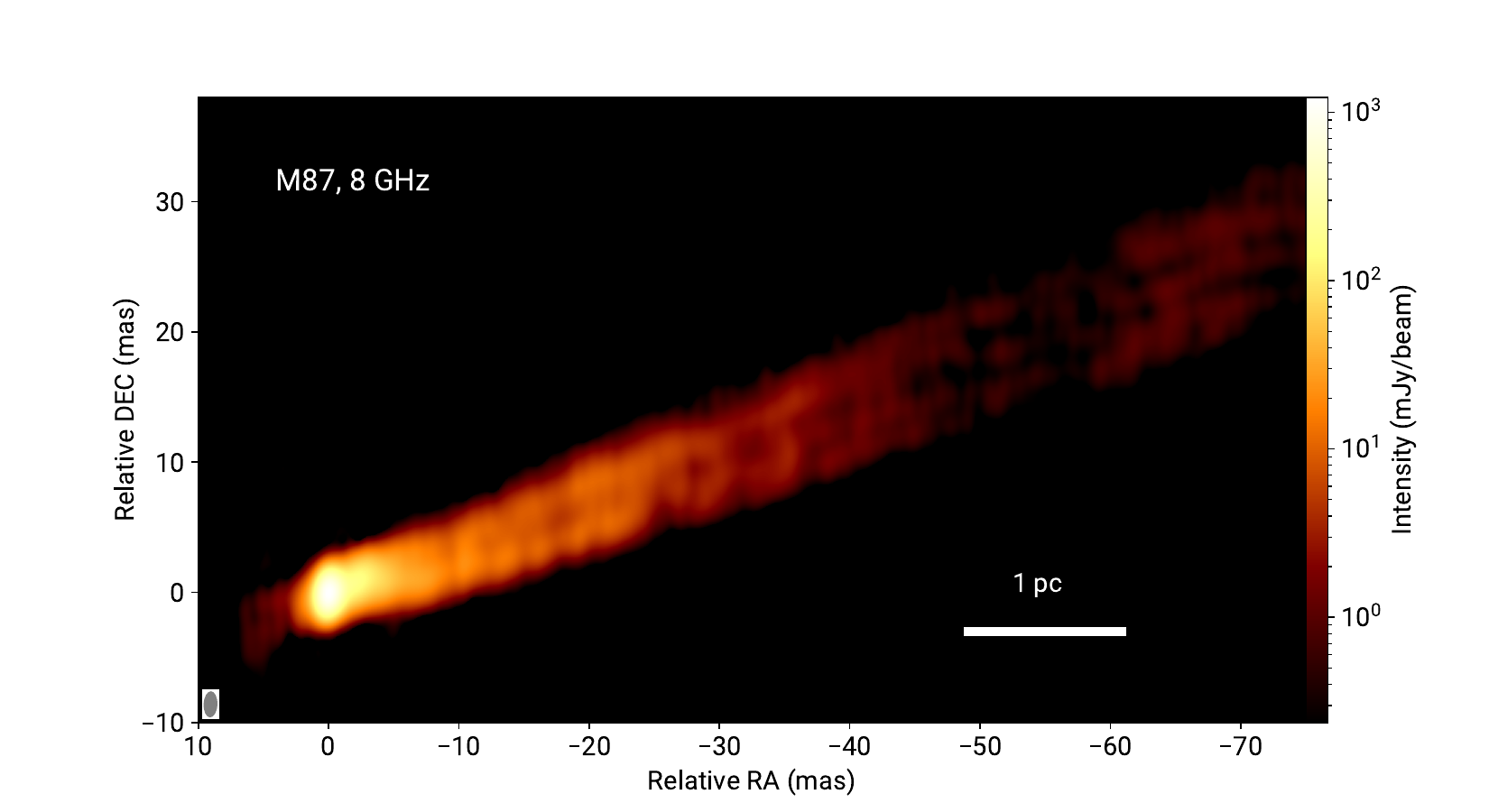}
    \caption{Total intensity $\texttt{CLEAN}$ images of the jet in M87 at 15\,GHz (top panel, combined data from May 22 and 24, 2009) and 8\,GHz (bottom panel, data from May 23, 2009) with VLBA+Y1 configuration. Both images are reconstructed with a natural weighting of the visibility data. The peak flux densities are 1.23\,Jy/beam and 1.22\,Jy/beam, at 15 and 8\,GHz, respectively. The elliptical restoring beams (full width at half maximum, FWHM) shown at the bottom left corner of each panel are $1.2\times0.6$\,mas, $\mathrm{PA}=-10^\circ$ at 15~GHz, $2\times1$\,mas, $\mathrm{PA}=-2^\circ$ at 8\,GHz.} 
    \label{fig:stokes_i_no_EB}
\end{figure*}

Full Stokes, full track dual-frequency observations of M87 were made in May 2009 using the Very Large Baseline Array (VLBA) involving the Karl Jansky Very Large Array (VLA) single antenna (Y1) and the 100-m Effelsberg Radio Telescope (Eb). The project code from the NRAO archive is BK145. We used Y1 to obtain the shortest baseline possible with this array, thereby improving the extended flux sensitivity, which is crucial in the case of the M87 jet. The Effelsberg was used to obtain higher sensitivity to weak and compact structures. 
Observations were performed on three consecutive days.
This included U-band with a central frequency 15.4\,GHz (later 15\,GHz) on 22 and 24 May as well as X-band with a central frequency 8.4\,GHz (later 8\,GHz) on 23 May. The length of each day's observation was 12 hours. The data were recorded in 16 baseband channels (intermediate frequencies, IFs), each of 8 MHz bandwidths using 512 Mbps recording rate and 2-bit sampling. Both right- and left-hand polarisations were recorded simultaneously, giving a total observing bandwidth of 32\,MHz in each polarisation.

\subsection{A priori data calibration}

The data were edited and calibrated following traditional methods\footnote{\url{http://www.aips.nrao.edu/CookHTML/CookBookch4.html}} in AIPS \citep{Greisen2003}. 
The quasar 0923+392 was used as a fringe finder.
Polarisation leakage was calibrated with the AIPS task LPCAL using a 1308+326 calibrator. The electric vector polarisation angle (EVPA) was calibrated by comparing the measured EVPAs of OJ~287, 0923+392 and 1308+326 with the values obtained from the UMRAO database\footnote{\url{https://dept.astro.lsa.umich.edu/datasets/umrao.php}} and the VLA polarisation database\footnote{\url{https://www.aoc.nrao.edu/~smyers/calibration/2009/}} at 8 and 15\,GHz. Due to a paucity of measurements, the values for the exact day of VLBA observations were found by linear interpolation of the UMRAO EVPAs from nearby dates. 
The final visibility data were averaged over 10 seconds.
The AIPS task UVMOD was applied to the 22 and 24 May visibility data sets, creating a combined $(u,v)$-file at 15\,GHz, to provide comparable to 8~GHz data sensitivity and, consequently, a higher quality of spectral index reconstruction. 

\subsection{Imaging}
\label{sec:imaging}

We used the Caltech DIFMAP \citep{Shepherd1994} software to obtain Stokes I, Q and U maps from the calibrated visibility data using hybrid mapping in combination with super-uniform, uniform and natural weighting with pixel sizes of 0.08~mas at 15~GHz and 0.16 mas at 8\,GHz. The final imaging results yielded >20,000:1 dynamic range at each frequency. The noise level is estimated  \citep{Hovatta2014} as $\sigma = ( \sigma_{\textrm{rms}}^2 + (1.5\sigma_{\textrm{rms}})^2 )^{1/2} \approx 1.8(\sigma_{\textrm{rms}})$, where the first term is connected with the off-source rms noise level of the image $\sigma_{\textrm{rms}}$ $\simeq$ 35 $\mu \textrm{Jy beam}^{-1}$, and the second term is connected with the uncertainties of the $\texttt{CLEAN}$ \citep{1980A&A....89..377C} procedure. 

We created two datasets: with and without data on baselines to Eb. 
All analysis in this paper was done with both datasets. For convenience, the final results in each section of the paper represent an average between results obtained with and without Eb baselines. In sections, where the difference between datasets is significant, the results of only one dataset are shown. For example, the resulting spectral index maps with the Effelsberg data show higher uncertainties as compared to the VLBA + Y1 configuration. Thus, the images presented in this paper use the conservative VLBA + Y1 configuration only as the ``Atlantic gap''  gives poor $(u,v)$-coverage between the main array and the Eb, which makes the image reconstruction process more complicated, which can be seen in the spectral index map. In addition, this makes a Kelvin--Helmholtz thread analysis more complicated. The resulting Stokes I images at 8 and 15\,GHz are shown in \autoref{fig:stokes_i_no_EB}.

The lack of short baselines leads to a significant loss of flux from faint and extended features far from the core, especially at relatively high frequencies such as 8 and 15\,GHz. Nevertheless, the data allowed reconstruction of the jet image until $\approx$~450\,mas and detect an HST-1 feature located at $\approx$~850\,mas from the core. To achieve a better dynamic range and to compare the results with the previous VLBA and EVN (European VLBI Network) HST-1 detections \citep{2007ApJ...663L..65C, 2010A&A...515A..38C, 2012A&A...538L..10G}, the images were convolved with an $8\times3$\,mas beam, corresponding to a VLBA beam at $\lambda$ 18\,cm (\autoref{fig:M87_full}).

\section{The M87 jet structure}
\label{sec:jet_structure}

\subsection{Jet shape}
\label{sec:jet_shape}

\begin{figure*}
    \centering
	\includegraphics[width=\linewidth, trim=1cm 3.4cm 1cm 4cm, clip]{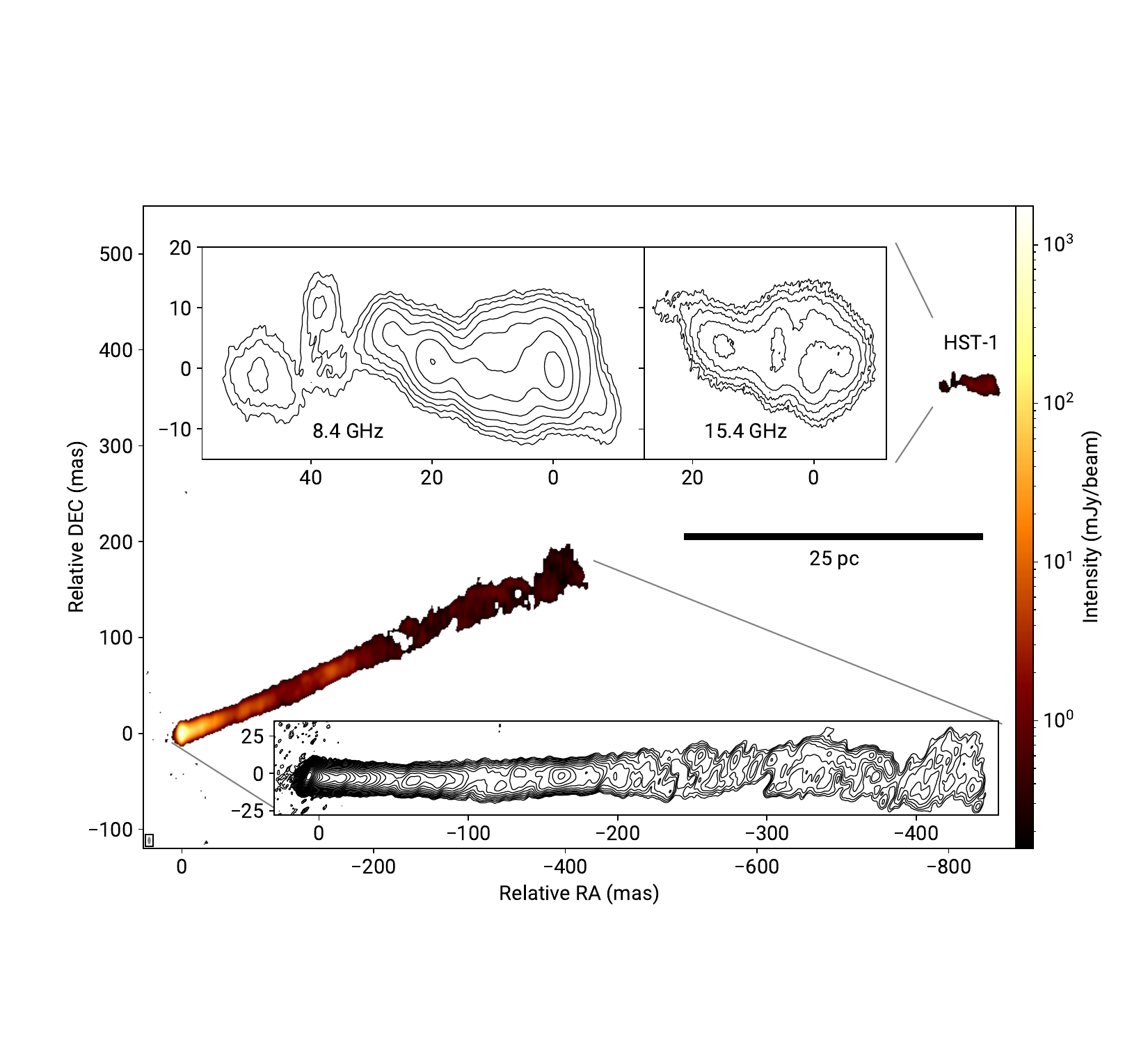}
    \caption{Total intensity image of the jet in M\,87 at 8 GHz (colour) restored with an elliptical beam of $8\times3$\,mas, $\mathrm{PA}=0^\circ$ approximately equivalent to the restoring beam of a VLBA observation of M\,87 at 18\,cm. The peak flux density in the image is 1.7\,Jy/beam. The HST-1 feature is located at $\approx 850$\,mas from the core.  The insets show the contour image of the inner 450\,mas of the jet at 15\,GHz (bottom) and the HST-1 region at 8\,GHz (top left) and 15\,GHz (top right), with the lowest contour level at 156\,$\mu$Jy/beam and successive contour levels increasing by a factor of $\sqrt{2}$. The HST-1 feature has a peak flux density of 1.4\,mJy/beam and 0.8\,mJy/beam at 8\,GHz and 15\,GHz, respectively.}
    \label{fig:M87_full}
\end{figure*}

Jet morphology provides crucial information for understanding the formation and propagation of relativistic outflows. The observed shape of an AGN jet depends on a number of parameters describing physical conditions in the jet plasma and in the ambient medium. So the knowledge of the shape and geometry of the flow can help us estimate intrinsic jet parameters \citep{2017MNRAS.468.4992P, 2017ApJ...834...65A, 2020MNRAS.495.3576K, 2019MNRAS.489.1197N, 2020MNRAS.498.2532N}. 

In M87, the jet manifests a complex and asymmetric triple-ridge internal structure developing in a predominantly straight, expanding flow. To analyse this structure, we first determine the direction of the jet axis by estimating the overall jet position angle (PA). To avoid potential uncertainties and errors due to the local curvature of the jet near its origin, we estimated the general PA of the jet position of the knot HST-1 in the large-scale image of the jet obtained from our data and presented in \autoref{fig:M87_full}. Using the intensity maximum of the HST-1, we obtained $\mathrm{P.A.} = 293{\fdg}3 \pm 0{\fdg}5$. Note the significant difference of several degrees from results by \citep{2022ApJS..260....4P} due to the different utilized methods of P.A.\ measurements.

\begin{figure}
    \centering
	\includegraphics[width=\columnwidth, trim=0.5cm 1.0cm 0.5cm 0cm, clip]{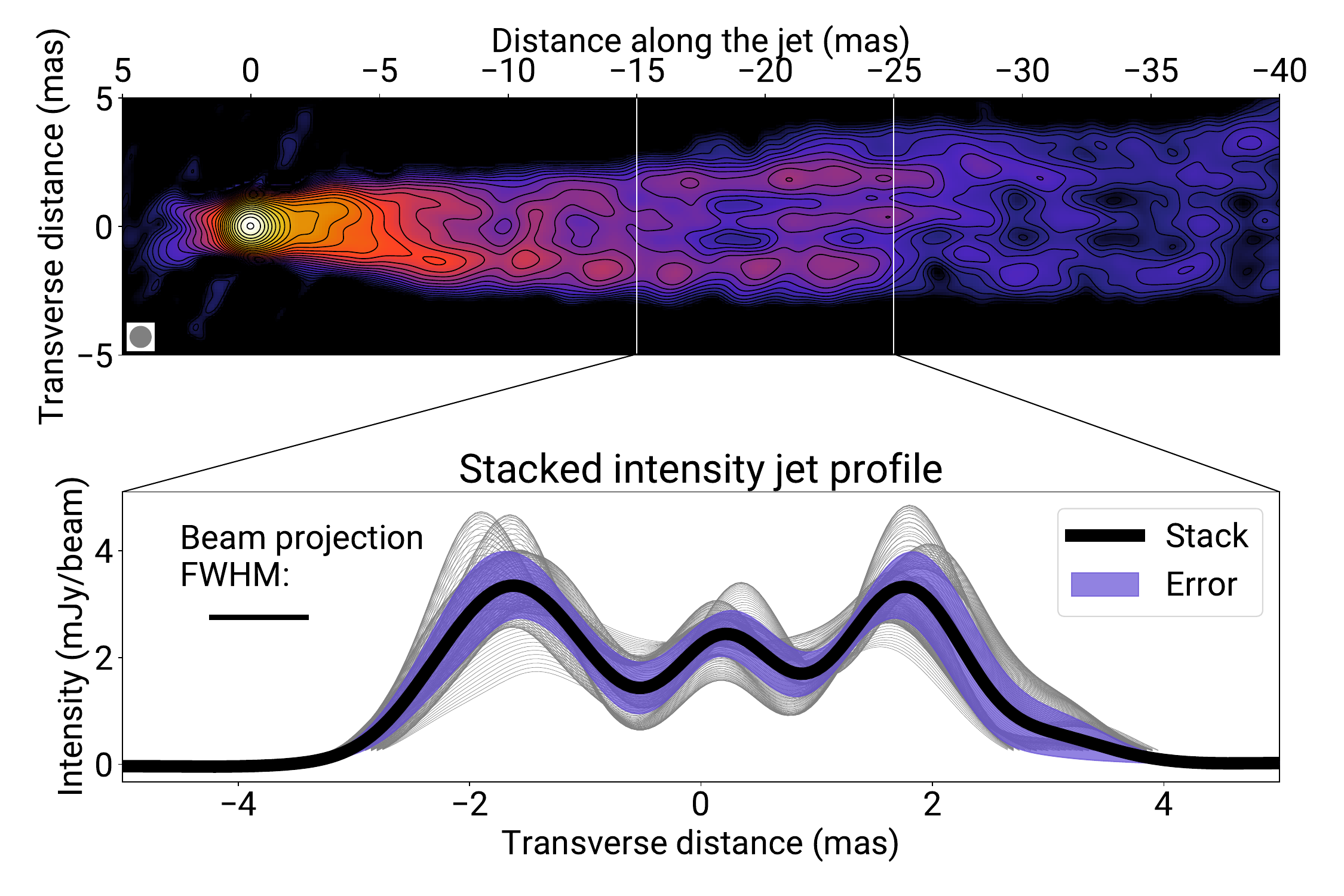}
    \caption{Top: Total intensity image of the jet in M87 at 15\,GHz, restored with a circular beam of 0.84\,mas FWHM (equivalent in area to the elliptical restoring beam used in \autoref{fig:stokes_i_no_EB}).
    The peak flux density is 1.23\,Jy/beam. The intensity contours start at 190\,$\mu$Jy/beam and successive contour levels increase by a factor of $\sqrt{2}$. Bottom: The stacked profile of the jet brightness obtained by averaging all individual transverse profiles measured with a step 0.05\,mas in the jet at 15--25\,mas separations from the core (solid black line) and its statistical error (violet filling).}
    \label{fig:shapshot}
\end{figure}

To simplify the further analysis, we rotate the jet images clockwise by $\psi = 23{\fdg}3$ to coincide the jet axis and relative right ascension and obtain transverse profiles of brightness distribution in the jet. In order to highlight the triple-ridge structure, a stack profile is shown in \autoref{fig:shapshot}. It was made by averaging all individual profiles with a step of 0.05 mas at 15--25 mas from the VLBI core. In the jet images produced with the nominal restoring beam, the jet structure up to $\approx 40$\,mas and $\approx 80$\,mas, in the 15\,GHz and 8\,GHz image, respectively can be traced. 
The 8\,GHz image can be further convolved with a larger beam $\approx$~1/3 of the jet width in the outer parts (3\,mas), which allows us to trace the jet structure with the transverse brightness distribution profile measured up to $\approx 200$\,mas separation from the jet origin.

Below, we apply the transverse profiles to measure the evolution of the jet width and quantify the properties of the internal structure of the flow. 
For both of these tasks, we fitted the obtained profiles with multiple Gaussian components defined as: 
\begin{equation}
    I(x) = \sum_i A_i e^{- (x_i-b_i) / 2c_i^2},
\label{eq:gaussian}
\end{equation}
where $x$ is a transverse distance, the $i$-th component is described by $A_i$ peak amplitude, $b_i$ peak location, and FWHM full width at half maximum, calculated from $\textrm{FWHM}_i = 2(2\ln 2 )^{1/2}c_i$.

The number of Gaussian components used in fitting of each profile was chosen using an $\chi^2$ analysis. Each profile was fitted by a multi-Gaussian function with a different number of components, where the maximum quantity is 3. Afterwards, the reduced $\chi^2$ with $q$ degrees of freedom for all fits for a given intensity profile were compared according to \cite{1976ApJ...210..642A}. The number of degrees of freedom $q = n - p$, where $n$ is the number of beams which can be fitted inside the transverse profile, and $p$ is the number of the fitting parameters. Using a cumulative distribution function of $\chi^{2}$, the theoretical value of $\Delta(q, \alpha)$ $ \geq $ $\chi^2$ can be calculated, where the probability of this expression equals $\alpha = 95\%$ (the confidence level). In this method, the $q+1$ degrees of freedom model is preferred if the following ratio is satisfied:
\begin{equation}
    \frac{\chi_{q}^{2}}{\chi_{q+1} ^{2}} \geq \frac{\Delta(q+1, \alpha)}{\Delta(q, \alpha)}.
\end{equation}

The width of the jet in a particular image profile was defined as a Full Width at a Quarter Maximum (FHQM) of the fit of the profile. The quarter maximum level was chosen because of the complex and asymmetric transverse structure of the M87 jet, where multiple components can have more than twice the brightness difference. In this case, the usage of the half maximum level can be misleading by measuring the width of one component instead of the whole jet.

\begin{figure}
    \centering
	\includegraphics[width=\columnwidth, trim=0cm 0.5cm 0cm 0cm]{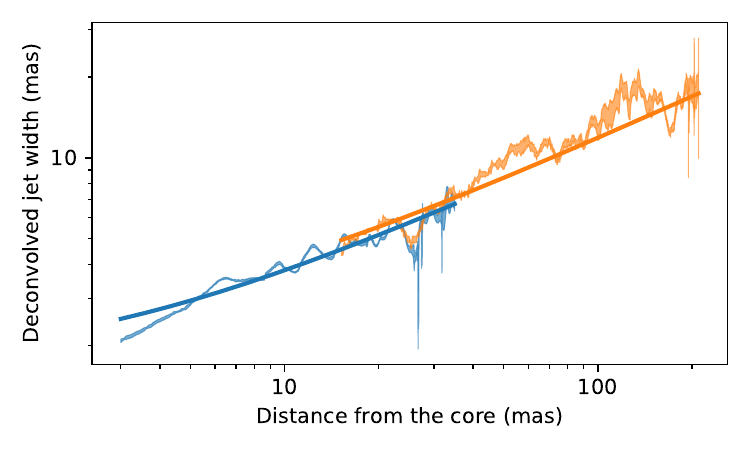}
    \caption{Expansion profile of the M87 jet at 15 and 8 GHz. The measurements of a jet width and uncertainties are shown as semi-transparent plots. The best-fit results are shown by solid lines. The blue colour represents the measurements taken with the 15\,GHz intensity model convolved with a circular beam of 0.86 mas FWHM which has an equivalent area of a 15\,GHz elliptical beam. The orange colour displays jet widths measured in the 8 GHz intensity model convolved with a 3 mas circular beam. This beam size was used to have the ability to trace extended up to 200\,mas faint jet, that is barely visible with the conservative beam (\autoref{sec:jet_shape}).  }
    \label{fig:geometry}
\end{figure}

\begin{figure*}
    \centering
	\includegraphics[width=\linewidth, trim=2cm 1.4cm 2.5cm 2cm, clip]{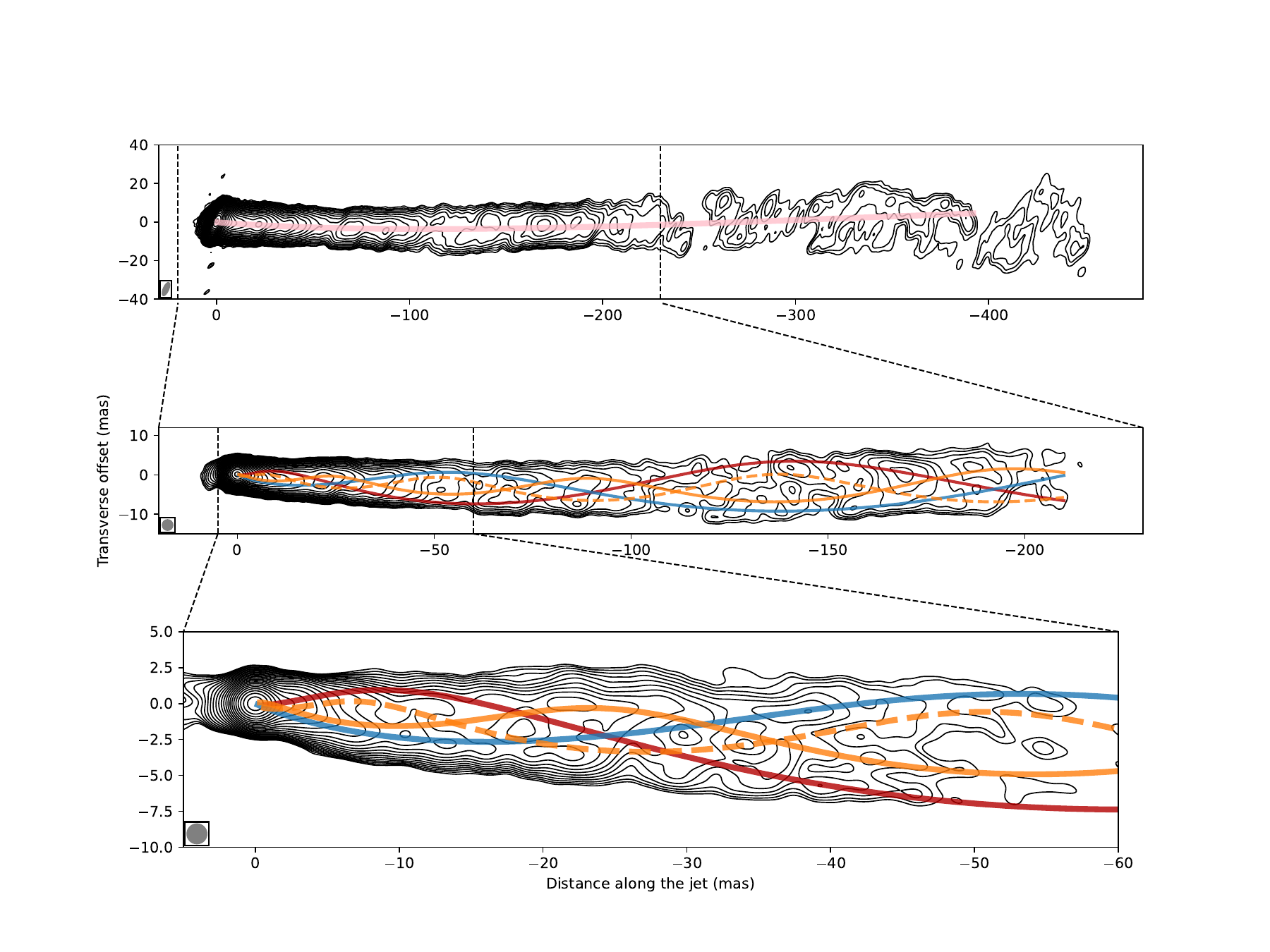}
    \caption{Modelling the transverse profiles of the jet brightness with oscillatory patterns. Semi-transparent curves represent the respective KH model. The dashed orange curve shows the model of the same colour but with the $180^{\circ}$ phase offset, indicating the approximate trajectory of the anticipated secondary thread of the elliptical body mode.  All Stokes I images shown in contours represent the same data restored with a different beam. Top panel: the beam ($8\times3$\,mas, $\mathrm{PA}=0^\circ$) is shown at the bottom left corner. The map peak flux density is 1.7\,Jy/beam. The intensity contours start at 260\,$\mu$Jy/beam level. Middle: the image is convolved with 3\,mas circular beam and has a peak flux density of 1.7\,Jy/beam. The intensity contours start at 360\,$\mu$Jy/beam level. Bottom: the innermost 60\,mas section of jet. The image is convolved with a circular beam of 1.56\,mas in diameter, which has an equivalent area of the elliptical beam used in \autoref{fig:stokes_i_no_EB}. The image peak flux density is 1.3\,Jy/beam, and the intensity contours start at 494~$\mu$Jy/beam level. In all panels, successive contour levels increase by a factor of $\sqrt{2}$. The jet images were rotated by the $\psi = 23{\fdg}3$, which was estimated in \autoref{sec:jet_shape}.}
    \label{fig:plasma_instab}
\end{figure*}

\begin{figure*}
    \centering
	\includegraphics[width=\linewidth, trim=0cm 0cm 0cm 0cm, clip]{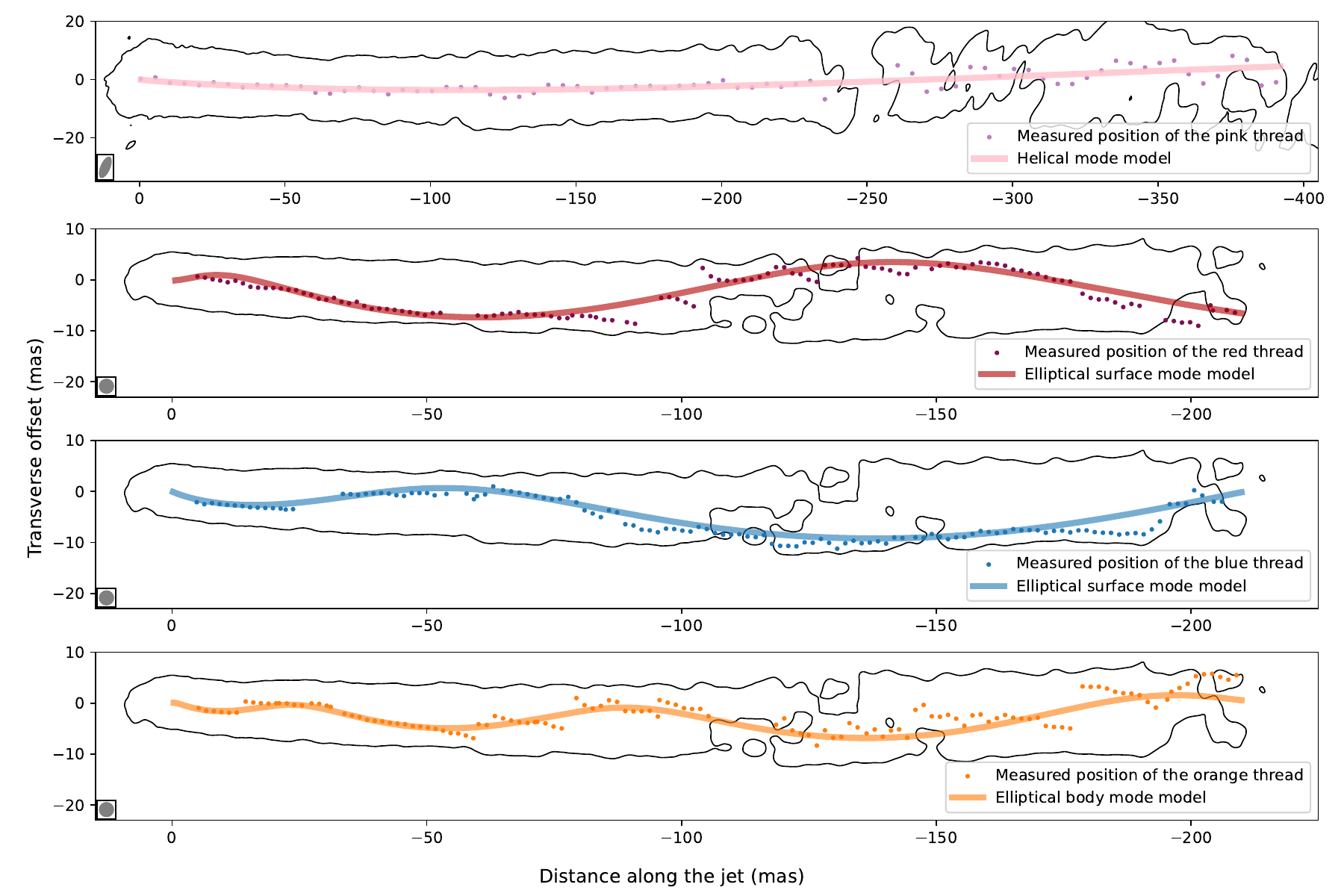}
    \caption{Jet decomposition by the oscillatory modes. Each extracted thread from the jet intensity images is presented in a separate sub-figure as a dotted plot with the corresponding colour according to the values listed in \autoref{tab:KH_fit}. To improve clarity, every tenth data point is plotted. A fitted KH model is displayed here as a curve with a corresponding colour. All plots show the lowest contour of the corresponding intensity image from \autoref{fig:plasma_instab}. Top panel displays the intensity contour at 260\,$\mu$Jy/beam level. The beam ($8\times3$\,mas, $\mathrm{PA}=0^\circ$) is indicated at the bottom left corner. The remaining sub-figures display intensity contour at the 360 $\mu$Jy/beam level, where the image was convolved with a 3 mas circular beam. The jet images were rotated by an angle of $\psi = 23{\fdg}3$, which was estimated in \autoref{sec:jet_shape}.}
    \label{fig:plasma_instab_fit}
\end{figure*}

After obtaining the jet width as described above, we deconvolved it with the beam projected onto a transverse profile. In our case, one-dimensional (1D) deconvolution procedure was done according to $w = (\textrm{FWQM}_\textrm{jet width}^2 - \textrm{FWQM}_\textrm{beam}^2)^{0.5}$, where $w$ is the deconvolved jet width. In this case, a 1D deconvolved profile can be distorted by a two-dimensional (2D) elliptical beam which influences several profiles asymmetrically. In our data, the major axis of a beam is oriented roughly perpendicular to the jet direction, thus the distortions should be negligible. At the same time, in addition to restoring elliptical beams, we reconstructed images using equivalent area circular beams. Such maps should deliver even less biased results and be easier to analyze. Indeed no significant difference was found between the geometry obtained with circular and elliptical beams. 

To investigate the jet geometry, we follow \cite {2020MNRAS.495.3576K} fitting the expansion profile by a power-law function $w\propto (r+r_0)^k$, where $r$ is the distance from the apparent VLBI core, $r_0$ is the distance from the core to the jet base, the expansion index $k$ shows the jet geometry. In \autoref{fig:geometry} the fitting results of the expansion profile are presented. For this, we used the jet images convolved with a circular beam of the FWHM-size 0.86 and 1.56 mas having an equivalent area of 15 and 8\,GHz elliptical beam, respectively. In addition, the geometry of the 200\,mas long jet image at 8\,GHz convolved with 3\,mas circular beam is also presented in \autoref{fig:geometry}. Jet geometry parameters were obtained for images with different array configurations (VLBA+Y1; VLBA+Y1+Eb), frequencies (8 and 15 GHz), and for extended 200\,mas jet at 8\,GHz. Finally, all obtained results were averaged to obtain the final M87 jet geometry parameters: $k = 0.532 \pm 0.008$, $r_0 = 3.8 \pm 1.1$~mas. The $r_0$ estimate by this method has a large uncertainty and is higher than expected; compare with \cite{Hada2011}, where the VLBI core separations are about 0.2~mas for 8 and 15\,GHz. 
A possible $r_0$ overestimation may result from methodological complications. First, the $r_0$ value in the fitting procedure is sensitive to the absolute value of the jet width in contrast to $k$, which is sensitive to the relative width. The uncertainty of absolute values is mainly driven by the deconvolution procedure. Second, physical conditions in the jet can change geometry locally, leading to over- or underestimations in $r_0$ estimates. In order to improve future measurements, a dedicated study of a possible deconvolution bias and geometry variations is needed on a large sample of AGN jets \citep[e.g.][]{2020MNRAS.495.3576K}.

\subsection{Kelvin-Helmholtz instability}
\label{sec:plasma_instability}

\begin{table}
    \caption{Identification of Kelvin-Helmholtz plasma instability modes.}
    \label{tab:KH_fit}
    \centering
    \begin{threeparttable}
        \begin{tabular}{lcccl}
            \hline
            Thread  & $A_0$ (mas)  & $\lambda_0$ (mas) & $\phi_0$ ($^{\circ}$) &  Mode \\
            (1) & (2) & (3) & (4) & (5)  \\
            \hline
            Pink & $0.42 \pm 0.06$ & $126 \pm 30$ & $188 \pm 18$  & H$_\mathrm{s}$ \\
            Red & $0.64 \pm 0.02$ & $33.2 \pm 1.4$ & $340 \pm 14$ & E$_\mathrm{s}$ \\
            Blue & $0.53 \pm 0.06$ & $32.7 \pm 1.2$ & $156 \pm 11$ &E$_\mathrm{s}$ \\
            Orange & $0.308 \pm 0.002$ & $16.8 \pm 0.2$ & $66 \pm 7$ & E$_\mathrm{b1}$ \\
            
            \hline
            
        \end{tabular}
        \begin{tablenotes}
            \item The table shows: 
            (1) thread name by colour according to \autoref{fig:plasma_instab}, (2) amplitude and (3) wavelength of an oscillatory pattern in R$_{\textrm{jet}} = 1$\,mas area, (4) phase of a helix, (5) identified instability modes: E$_\mathrm{s}$ (Elliptical surface mode), E$_\mathrm{b1}$ (First-order elliptical body mode), H$_\mathrm{s}$ (Helical surface mode).
        \end{tablenotes}
        
    \end{threeparttable}

\end{table}

Transverse oscillations in the M87 jet were analysed in previous studies. The 17-year observations at 43\,GHz by VLBA show a shift of the transverse position of the jet on a quasi-periodic 10-year timescale that is consistent with the Kelvin-Helmholtz instability \citep{Walker}. Recent studies of the M87 jet show a 1-year period wiggles in multi-epoch 22\,GHz KaVA VLBI observations \citep{2023Galax..11...33R}. It is still unclear what is the origin of these fast oscillations, but current-driven instability (CDI) was chosen as a preferable mechanism. \citet{2023NaturePortfolio_Yuzhu} analysed the periodicity of the P.A. of the jet using 22-43\,GHz EAVN, VLBA and EATING VLBI observations, where the phenomenon was interpreted as the jet nozzle precession with a period of $\approx 11$\,years. 

The data analysed in this article provide the most prominent, sensitive and extended pc-scale jet structure. In \autoref{fig:stokes_i_no_EB} the brighter jet limb changes from the northern to the southern edge of the jet at a relative RA $\approx -10$\,mas and vice versa at a relative RA $\approx -25$\,mas), and this effect is observed in both the 8\,GHz and the 15\,GHz maps (Figures~\ref{fig:stokes_i_no_EB}--\ref{fig:shapshot}). Similar behaviour (changes of the brighter limb and multiple threads inside the jet) can be seen in the pc scales VLBA and RadioAstron observations \citep{Walker, 2021cosp...43E1398S}, the kpc scales in the VLA \citep{1989ApJ...340..698O, Pasetto_2021} and the HST \citep{1996ApJ...473..254S} images. The pattern was interpreted as resulting from the KH instability in the flow \citep{2003NewAR..47..629L}. 

At the distances within 15-25\,mas from the core a triple-ridge structure is observed in 15\,GHz map (Figures~\ref{fig:stokes_i_no_EB}, \ref{fig:shapshot}). The central filament has been observed before in several studies both in proximity to the radio core \citep{2016ApJ...833...56A, Walker, 2018A&A...616A.188K} and downstream \citep{2017Galax...5....2H} and associated with the spine in the `spine-sheath' jet model \citep{2016AA...595A..54M}. The simulations done by \cite{2023arXiv230112861P} show that the central filament might be found in an edge-brightened model due to a $\texttt{CLEAN}$ imaging artefact. At 15\,GHz the jet width coincides with three beam sizes in that region. This can cause an overlay of beam sidelobes, creating the apparent central filament. This hypothesis is supported by the absence of a prominent central ridge at 8\,GHz image, where the jet width is about two beams. For this purpose we use 8\,GHz maps for this analysis. 

In \autoref{sec:jet_shape}, the jet transverse profiles were modelled by Gaussian components. The positions of the peaks of those components obtained for the inner and outer parts of the jet were superimposed together on the intensity image (bottom plots of \autoref{fig:plasma_instab} ), thus allowing us to trace the development of the jet structure on scales of up to $\sim$200\,mas from the observed jet origin. The revealed jet pattern suggests the presence of up to three intertwining helical threads inside the jet, which can be related to the development of KH instability inferred for the kpc-scale jet. 

Identifications of each of the individual threads were made by requiring a continuous and smooth evolution of thread parameters, such as position, intensity, and width along the jet. Using this approach, a self-consistent picture of the evolving thread-like patterns inside the jet was reconstructed.

In addition to the rich evolution of the internal structure, the jet exhibits a slight bend  $\approx$80\,mas which is clearly visible in \autoref{fig:M87_full}. To quantify the position and the magnitude of this bent, the 400\,mas jet image with 18\,cm VLBA beam was used. The image was rotated $23{\fdg}3$ clockwise, and all transverse intensity profiles were fitted by a single Gaussian, obtaining the ridgeline (the blue line in the top of \autoref{fig:plasma_instab}).

Both, the observed bent and the pronounced internal structure observed in the jet can result from the development of the KH instability in the flow \citep{Hardee_2003,2003NewAR..47..629L}. The changes in the ridge line and the evolution of the individual threads identified inside the flow can be well described by a three-dimensional (3D) helix in a Cartesian coordinate system ($x, y, z$):
\begin{equation}
    [x,y] = A[\sin,\cos]{((2 \pi / \lambda )z + \phi )},
    \label{eq:helix}
\end{equation}
where $z$ is the distance from the jet origin, $\phi$ is the phase, and the amplitude and the wavelength depend on the jet radius $A \propto R_{\textrm{jet}(z)}$, $\lambda \propto R_{\textrm{jet}(z)}$ \citep{2000ApJ...533..176H}. 

First, we fitted the global jet curvature by a 3D helix projected onto the sky plane, for which the jet angle to the line of sight $\theta = 17{\fdg}2$ (\citet{2016AA...595A..54M}) was used. The resultant mode can be associated with the helical surface mode (H$_\mathrm{s}$) of the KH instability as it apparently leads to oscillations of the entire jet around its average propagation direction. In order to measure more accurately the parameters of the modes affecting the internal structure of the flow, this mode was subtracted from the component positions before the fitting procedure. A robust identification of the other modes can be obtained
using the characteristic wavelength $ \lambda^{*} = \lambda_{nm} (n + 2m + 0.5)$, where $\lambda^{*}$ is the characteristic wavelength, $\lambda_i$ is the observed wavelength, $n$ is the azimuthal wavenumber and $m$ is the order of the mode. The pinch ($n = 0$), helical ($n = 1$) and elliptical ($n = 2$) modes are expected to be most prominent in relativistic jets. The order of the mode $m$ determines whether the corresponding perturbation affects the surface ($m$ = 0) or the interior ($m > 0$) of the jet. The characteristic wavelength depends only on the physical conditions in the jet; thus, should have a similar value for different modes \citep{doi:10.1126/science.1063239}. The final fitting results are presented in \autoref{fig:plasma_instab} and \autoref{fig:plasma_instab_fit} by semi-transparent thick pink, red, orange and blue curves. A zoom into the innermost 60\,mas section of the jet is shown in the bottom panel of \autoref{fig:plasma_instab}. The identification criteria discussed earlier were applied to the extracted individual threads resulting in one plausible combination of KH modes shown in \autoref{tab:KH_fit}. The red and blue threads have nearly equal wavelengths and $\approx$180$^{\circ}$ phase difference, which is expected to result from the two regions of increased pressure and density produced in the jet by the elliptical surface (E$_\mathrm{s}$) mode of instability \citep{2000ApJ...533..176H}. The orange thread has a much shorter wavelength with a smaller amplitude, which is expected in the case of the body mode. In the case of a first-order elliptical body mode (E$_\mathrm{b1}$), one should observe the accompanied thread. For illustration purposes, the thread with the same parameters, but with the 180$^{\circ}$ phase difference was plotted as a dashed curve of the same colour in \autoref{fig:plasma_instab}. Due to the lack of high-enough resolution, the accompanied thread was not identified robustly and extracted from the data since it is blended with the other modes. However, it is observed that the dashed curve plotted in \autoref{fig:plasma_instab} complements the KH model well, describing the intensity images better. 

To verify the robustness of the fitting procedure, we generated a synthetic dataset reproducing the internal structure with the parameters of all fitted modes. For this, we first created a model image using the parabolic jet model threaded by the KH modes obtained from the fits to the 8\,GHz image. Then we created artificial $uv$-data using the same $uv$-coverage and thermal noise, as in the original 8\,GHz data set. The same algorithms of the image reconstruction and the KH modes fitting were applied to the artificial data set. The derived parameters of the KH modes were found to be consistent with the ground truth values within 25\%. First, this demonstrates that our approach can reconstruct the parameters with the accuracy required to distinguish between two modes (surface and body). Second, the obtained accuracy of the method shows that two threads of the elliptical surface mode are consistent with having the same observed wavelength and the 180$^\circ$ phase offset.

\section{Spectral index map}
\label{sec:sp_ind}

\begin{figure*}
    \centering
    \includegraphics[width=\linewidth,trim=2.5cm 1.2cm 1.2cm 1.5cm, clip]{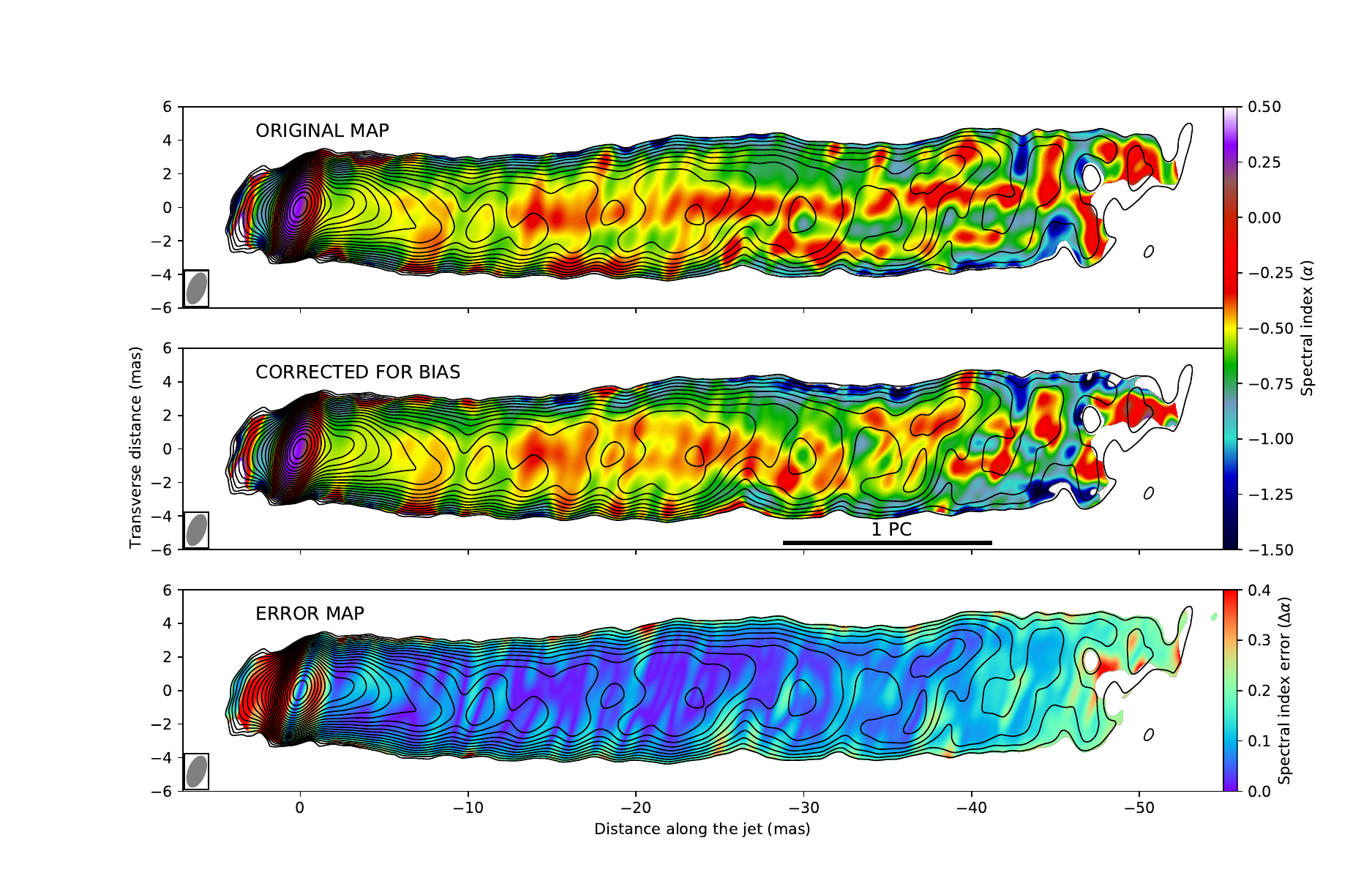}

    \caption{Spectral index map between 8 and 15\,GHz, shown as a false-colour image before correction (top), after correction (middle) and error (bottom) map. All images are rotated 19$^{\circ}$ clockwise. The rotation of the map applied here does not correspond to the global jet PA to follow the local curvature of the jet. 
    The size of the common 8 GHz restoring beam is displayed at the bottom left corner and is equivalent to $2\times1$~mas at $\mathrm{PA}=-2^\circ$ ellipse. The contours represent the 8\,GHz total intensity map, starting from  360~$\mu$Jy/beam level and increasing by a factor of $\sqrt{2}$. Only the inner 50 mas of the jet is shown due to the large spectral index errors in the outer regions. }
    \label{fig:sp_map_noEB_max}
\end{figure*}

The main mechanism of AGN radio jet emission is synchrotron radiation.
Assuming a power-law particle energy distribution $N(E)dE \propto E^{-p}dE$, the jet spectrum for the optically thin regions is then described by a distribution $I_{\nu} \propto \nu^{\alpha}$ with a spectral index $\alpha = (1-p)/2$. Therefore, the spectral index distribution provides important information about the physical conditions of different jet regions. 
Due to phase self-calibration, the information about absolute celestial coordinates is lost. Thus, images at different frequencies need to be well aligned to determine spectral index distributions. For this reason, the normalised 2D cross-correlation (2DCC) method was used \citep{Lewis1995, 2000ApJ...530..233W,  2008MNRAS.386..619C, 2013A&A...557A.105F, Hovatta2014}. The method uses optically thin parts of a jet as a reference. Due to their transparency, these parts can be assumed to be located in the same place at both frequencies. So choosing the jet image regions far away from the core and applying a normalized 2DCC, we can align the images at different frequencies.

\begin{figure}
    \centering
	\includegraphics[width=\columnwidth, trim=0.5cm 0.5cm 0.5cm 0.5cm, clip]{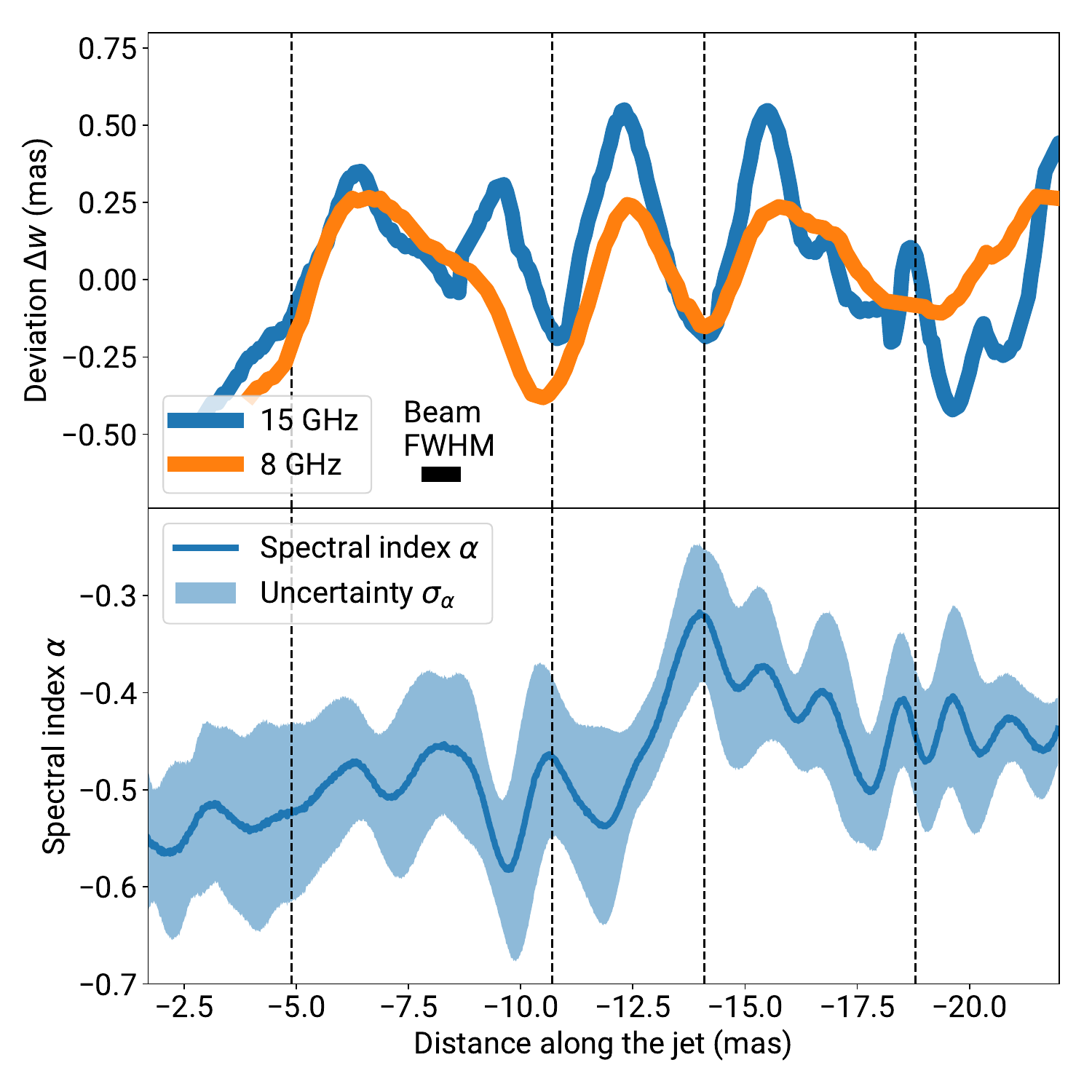}
    \caption{Deviations $\Delta w = w_{i} - w(r)_{\textrm{model}}$ between the measured jet widths $w_{i}$ and the fit by a power law curve $w(r)_{\textrm{model}} \propto (r + r_{0})^k$ (top). The blue lines represent 15\,GHz data, the orange lines represent the 8\,GHz data. The spectral index longitudinal profile is obtained from  \autoref{fig:sp_map_noEB_max} and presented here as a light-blue filled plot (bottom). 
    The plot shows quasi-periodic oscillations at both frequencies indicated by the black dashed lines.}
    \label{fig:shocks}
\end{figure}

The spectral index of the M87 jet obtained between 8 and 15\,GHz is shown at the top of \autoref{fig:sp_map_noEB_max}. Note that the two used images have comparable sensitivity. Due to the sparsity of the visibility plane coverage, a bias in the resulting spectral index image might be significant; see analysis in \citet{2023arXiv230112861P}. Thus, we estimated the effect of the bias and corrected for it, see details in \autoref{sec:bias_corr}. The final unbiased spectral index maps are shown in Figures~\ref{fig:sp_map_noEB_max} and \ref{fig:sp_map_noEB_max_circ}. The spectral index profile along the jet axis is presented in \autoref{fig:shocks}. 

Quasi-periodic oscillations of width along the jet are observed in the geometry profile \autoref{fig:geometry}. Additionally, \autoref{fig:shocks} shows deviations of a measured jet width at 8 and 15\,GHz from the fitted 15\,GHz power-law model. The oscillations exhibit quasi-periodic deviations from the parabolic shape with an amplitude up to 0.7\,mas. This result cannot be an imaging effect since the oscillations are seen at both observing frequencies with equal periodicity and at the same distances. Moreover, the contraction near 5 mas is also observed at different epochs \citep[][and Kravchenko et al. {\em in prep.}]{2016AA...595A..54M, Walker}. We suggest that this is caused by stationary recollimation shocks. The spectral index map profile in \autoref{fig:shocks} supports this assumption. There is evidence that recollimation shocks are seen in VLBI observations. In addition, a simulated VLBI total intensity map obtained by computing the radio continuum synchrotron emission using the relativistic magnetohydrodynamic (RMHD) model shows similar periodic contractions \citep{Gomez2016}. Periodic oscillations are also seen in other RMHD simulations \citep{2018ApJ...860..121F, Mizuno_2015}. Our results are qualitatively consistent with other observations and simulations, but individual simulations are needed for a detailed comparison. 

The M87 jet spectral index analysis results at 8--15\,GHz obtained by \citet{Hovatta2014} show a moderate steepening of the spectra within 10\,mas down to $\alpha \approx -1.5$. The recent results, observed at 22--43\,GHZ show a rapid steepening down to $\alpha \approx -2.5$ within 10\,mas from the VLBI core \citep{2023arXiv230301014R}. However, in this paper, there is no significant steepening in the same region. The interesting part of all these data is they all have a similar resolution, thus it is not a consequence of a resolution effect. In order to explain this discrepancy, several reasons are proposed. If the effect is intrinsic, then it can be caused by synchrotron radiation losses \citep{kardashev1962, pacholczyk1970}. This interpretation can be used in the case of NGC~315, where a similar effect was observed in two papers \citet{park2021jet} and \citet{2022A&A...664A.166R}. For this source in the same region, the spectral index steepens with the frequency. The other reasonable interpretation is the temporal variations of the spectral index in combination with the multi-layer jet structure. The last can produce several electron plasma populations with different energy distributions that will create a broken power-law spectrum. If the effect is instrumental, it can be caused by high-frequency flux losses due to $\texttt{CLEAN}$ bias \citep{2023arXiv230112861P}. In this case, the residuals of the high-frequency image are convolved with a high-frequency beam, that is smaller than the lower-frequency beam. Thus, if $\texttt{CLEAN}$ is not deep enough, this will reduce the flux value in the extended regions of the high-frequency image. In this paper, to avoid this effect, deep $\texttt{CLEAN}$ was performed. In addition, the unique feature of the data in this paper is twice the difference in exposure time between 8 and 15\,GHz. It was done to reduce the noise in 15\,GHz data to make it correspond to the 8\,GHz data. Thus, the spectral index map produced in this paper may be less affected by this effect, which is observed as the absence of the significant spectral index steepening along the jet.

\subsection{Core shift}

Due to synchrotron self-absorption, the location of the VLBI core depends on the frequency of observations as $\vec{r}_{c}(\nu) \propto \nu^{-1}$ (the so-called `core shift' effect), under the assumption that the jet is expanding freely, and there is equipartition between the particle kinetic and the magnetic field energy \citep{1998A&A...330...79L}. The observations show that the core shift in the many AGN jets indeed follows dependence $\vec{r}_{c}(\nu) \propto \nu^{-1}$ \citep{2011A&A...532A..38S}.
Moreover, using the phase-referencing multi-frequency observations of M87, \cite{2011Natur.477..185H} show that the core shift is described by $\vec{r}_{c}(\nu) \propto \nu^{-0.94 \pm 0.09}$.
Assuming $\vec{r}_{c}(\nu) \propto \nu^{-1}$, we can estimate the distance from the 15\,GHz core $\nu_{15}$ to the jet origin as:
\begin{equation}
    \vec{r_{c}}(\nu_{15}) = \frac{\Delta \vec{r}}{\frac{\nu_{15}}{\nu_{8}} - 1},
\end{equation}
where $\Delta \vec{r}$ is the core shift vector. To obtain the core shift vector, we measured the core positions and the alignment shift. To find the core positions, we used the MODELFIT function in DIFMAP: the M87 image was fitted with a model consisting of 2D Gaussian components. The coordinates of the model components, which have the highest brightness, were used as the core position relative to the phase centre ($\vec{r}_{8}$ for 8\,GHz and $\vec{r}_{15}$ for 15~GHz). The shift between images $\vec{r}_{8 \xrightarrow{} 15}$ was obtained using 2DCC at the beginning of \autoref{sec:sp_ind}, where images were aligned for constructing the spectral index map. Finally, we calculate the core shift vector as $\Delta \vec{r}_{8 \xrightarrow{} 15} = \vec{r}_{8} - \vec{r}_{15} - \vec{r}_{8 \xrightarrow{} 15}$. 

Averaging the results of the measurements for all restored images with different beams and antenna configurations, we estimated $|\vec{r}_{c}(\nu_{15})| = 0.2 \pm 0.1$ mas. This result is consistent with the phase-referencing core shift measurements \citet{Hada2011}.
Neglecting the distance from the central SMBH to the jet origin (which is predicted to be about 2.5--4\,$R_{\rm g}$ \citep{2012Sci...338..355D}), $|\vec{r}_{c}(\nu_{15})|$ represents the separation of the 15\,GHz VLBI core from the central SMBH. Finally, using $\theta = 17{\fdg}2 \pm 3{\fdg}3$ \citep{2016AA...595A..54M} (\autoref{sec:jet_to_counter-jet_flux_ratio}), we obtained the deprojected distance from the 15 GHz VLBI core to the black hole:  $r_{\textrm{deprojected}} = 0.7 \pm 0.3$ mas or $(5 \pm 2) \times 10^{-2}$ pc. 

\subsection{UNCERTAINTIES}\label{sec:BIAS}

\subsubsection{Statistical error of the map}\label{sec:statistic_and_systematic_erros}

The lower the intensity, the lower the signal-to-noise ratio and the larger the errors in the resulting spectral index map. The errors can also be influenced by uncertainties of image shift correction. To estimate the accuracy of the spectral index measurement, a spectral index error map was made. The error consists of the error caused by the uncertainty of the intensity measurement $\sigma_{\textrm{I}}$ and the uncertainty of the alignment of the image $\sigma_{\textrm{cs}}$, thus the spectral index random error is $\sigma_{\alpha} = (\sigma^{2}_{\textrm{I}} + \sigma^{2}_{\textrm{cs}} )^{1/2}$. 

For the errors caused by the intensity measurement uncertainties:
\begin{equation}
    \sigma_\mathrm{I} = \frac{1}{\log(\nu_\mathrm{U}/\nu_\mathrm{X})} \sqrt{{ \left( \frac{\sigma_\mathrm{X}}{I_\mathrm{X}} \right) } ^2+{ \left( \frac{\sigma_\mathrm{U}}{I_\mathrm{U}} \right)}^2}, 
\end{equation}
where $\sigma_\mathrm{X},\sigma_\mathrm{U}$ are the intensity maps errors,   and $I_\mathrm{X}$, $I_\mathrm{U}$ are the intensity maps at 8 and 15\,GHz, respectively. The errors caused by the uncertainty of the image alignment $\sigma_{\textrm{cs}}$ were obtained with the help of the algorithm:
\begin{enumerate}

    \item Define the uncertainty of alignment. In this case, the error is 2 pixels or 0.025 mas. 
    \item Shift the map in the alignment uncertainty value in four different directions and obtain spectral index maps.
    \item Subtract modulo for each of the four obtained maps with the original spectral index map.
    \item Average all four maps. 

\end{enumerate}

The bottom image of \autoref{fig:sp_map_noEB_max} displays the final spectral index error map. The error rises to the edges discussed earlier. Also, one can see that the error grows dramatically near the VLBI core. This is caused by the large gradient in intensity images near the core region, so the image alignment uncertainties produce significant errors. If the difference between neighbouring pixels along the jet axis in intensity images is high, then even a 1-pixel shift can dramatically change the resulting spectral index value. The rapid decrease of intensity in images southeast of the core is caused by the synchrotron self-absorption of a jet base and the transition from Doppler boosted to deboosted emission of the jet and the counter-jet. The region of a relatively large error ends at $\approx$ 0.5 mas southeast of the core, and the counter-jet area errors are significantly lower. This is another indication that the faint feature southeast to the core is counter-jet. The errors in the regions further than $\approx 50$ mas are high, so we decided to show a spectral index map within 50 mas.

\subsubsection{Effects of the sparsity of visibility plane}
\label{sec:convolution_effects_modelling}
Since the $(u,v)$-coverage is sparse, the restored beam takes a complicated shape with notable side lobes. The $\texttt{CLEAN}$ algorithm is unable to completely remove them because the flux is subtracted not only from the regions of maximum intensity but elsewhere. This creates a bias on a residual image, and artefacts can appear. After that, a $\texttt{CLEAN}$ model is convolved with a $\texttt{CLEAN}$ beam or a 2D Gaussian function approximation of a dirty beam. Thus, the side lobes are not taken into account, and convolution errors (bias) can appear. A spectral index map is especially sensitive to this effect. 

The coverage of the visibility plane depends on frequency. Even in the case of the same antennae configuration in dual-frequency observations, the pattern of the obtained $(u,v)$ coverage will be the same, but the size will be different. So the coverage at 8 GHz will look like a stretched version of the 15 GHz $(u,v)$ data. Thus, initially, the $(u,v)$-ranges do not correspond to each other, and it can cause errors or imaging artefacts which can become significant in spectral index maps, such as a steepening of the spectral index along a jet \citep{Hovatta2014}. 

To test how inconsistency of the $(u,v)$-ranges affects the spectral index map, the $(u,v)$-coverage was clipped to be the same in both bands. Surprisingly, a comparison of original and $(u,v)$-clipped data made an insignificant difference. Thus, $uv$-clipping was not applied to obtain the final spectral index map, shown in the paper. 
The sparsity effects are much more difficult to check since it is based on the fundamental problem of deconvolution. However, it is possible to estimate the influence of this effect on the spectral index map. Here are the steps of bias checking applied in the paper:

\begin{enumerate}
    \item Create an artificial model out of the $\texttt{CLEAN}$ components whose structure is similar to the real jet in its form and intensity.
    \item Create visibility functions with the same model but different $(u,v)$-coverage (at 8 and 15 GHz), using the UVMOD task in AIPS. 
    \item Do imaging in DIFMAP.
    \item Obtain a spectral index map.
\end{enumerate}

The obtained spectral index map should have zero values in a range of all maps, in the absence of a bias. In the real-intensity image, we can see the edge-brightening effect. To describe the jet and create a synthetic model, we assumed that the jet is hollow and used simple geometrical shapes. 
Since the edge-brightened jet is transparent and the core is opaque, we can define the core as a circle and the limbs as two expanding rails. The overall jet model was also inclined in a corresponding P.A.
This model was chosen to bring the simulation closer to the observations.

Analysing the results with different beams, we conclude that the spectral index map convolved with the 8\,GHz elliptical beam has minimal influence on the convolution effect with the error $\Delta \alpha \lessapprox$ 0.13 in the central ridge region. Since, in the case of the VLBA+Y1+Eb dataset, the long-baseline coverage is poor, the problem of the convolution effect is more significant compared to the data with flagged Eb.
It is seen that with increasing the resolution, the bias amplitude increases too. But its pattern is not truly aligned to the flattened spectrum in a ridge which was noticed in the spectral index map (\autoref{fig:sp_map_noEB_max}). The results of modelling also showed that the amplitude of the bias is less than the flat spectrum region values. Thereby, according to the analysis done above, the convolution effects do not affect the spectral index map. 

\subsubsection{Spectral index within the inner 6 mas}
\label{sec:evk}

To examine in detail the spectral index distribution in the inner 6~mas we used a circular beam obtained from an average of 8 and 15\,GHz elliptical beams equivalent area.
The resultant image revealed a double structure. The flattening of the spectral index in the image coincides with the peaks of total intensity at 8 and 15\,GHz, while the spectral index steepens toward the jet edges. 

We also made use of recent VLBA observations performed on 2018 April 28 simultaneously at 24 and 43\,GHz, which are presented in \citet{2020A&A...637L...6K}. The full-intensity images at two frequencies were convolved with a common 24\,GHz equal-area circular beam and were aligned using a 2DCC procedure. Besides, the $(u,v)$-coverage was matched for this pair of frequencies. 
The resultant spectral index map revealed a two-humped structure which corresponds well to the result at 8-15\,GHz observations. 

These two independent maps reveal a similar structure, see for details Figure~1 in \citet{2023arXiv230112861P}, but is it real? A detailed analysis was performed to check its significance and concluded that 
the two-humped structure is actually a product of an imaging bias. Due to its importance, the analysis was presented in a dedicated paper \cite{2023arXiv230112861P}.

\subsubsection{Bias correction of the spectral index images}
\label{sec:bias_corr}

\cite{2023arXiv230112861P} employed a series of simulations with various jet brightness models and dual-frequency VLBI data sets. They found that the spectral index maps of the M87 parsec scale jet are heavily affected by systematical effects. For the data set analysed in this paper, these effects flatten the spectrum in a series of stripes along the jet. This is similar to the observed image: two stripes near the core turn into the central and two outer stripes of the spectral flattening further out at $r \approx 10$ mas. The outer stripes are shown at the top of \autoref{fig:sp_map_noEB_max} by red horizontal stripes at 3 and $-3$ mas from the central stripe. The simulations of the data set reveal that the systematic spectral index effects trace the bias of the low frequency (8\,GHz) Stokes $I$ image. At the same time, the bias of the 15\,GHz Stokes $I$ image is down-weighted by convolving with a lower frequency $\texttt{CLEAN}$ beam. This made it possible to successfully compensate the spectral index bias in simulations by re-creating such effects in the high frequency (15\,GHz) Stokes $I$ image.
Indeed, assuming that the Stokes $I$ bias is small and does not heavily depend on the brightness distribution, which follows from the simulations, it cancels out in the expression for the spectral index.
The procedure of the spectral index bias correction consists of three steps. First is an interpolation of the original $\texttt{CLEAN}$ model at the 15\,GHz on the $uv$-points of the 8 GHz data set, creating visibility data of the 15\,GHz model with 8\,GHz $(u,v)$-coverage. Next, the resulting data set is imaged in the same way as the original 8\,GHz data set. In the final step, a bias-compensated spectral index map is produced (\autoref{fig:sp_map_noEB_max}). These results heavily contradict previous conclusions. It is evident, that simple intensity models, as they are used in \autoref{sec:convolution_effects_modelling}, are insufficient, and a more sophisticated approach is required to analyse the convolution effects and correct the bias in the spectral index map. 

\begin{figure*}
    \centering
	\includegraphics[width=0.48\linewidth, trim=4cm 0 5cm 0, clip]{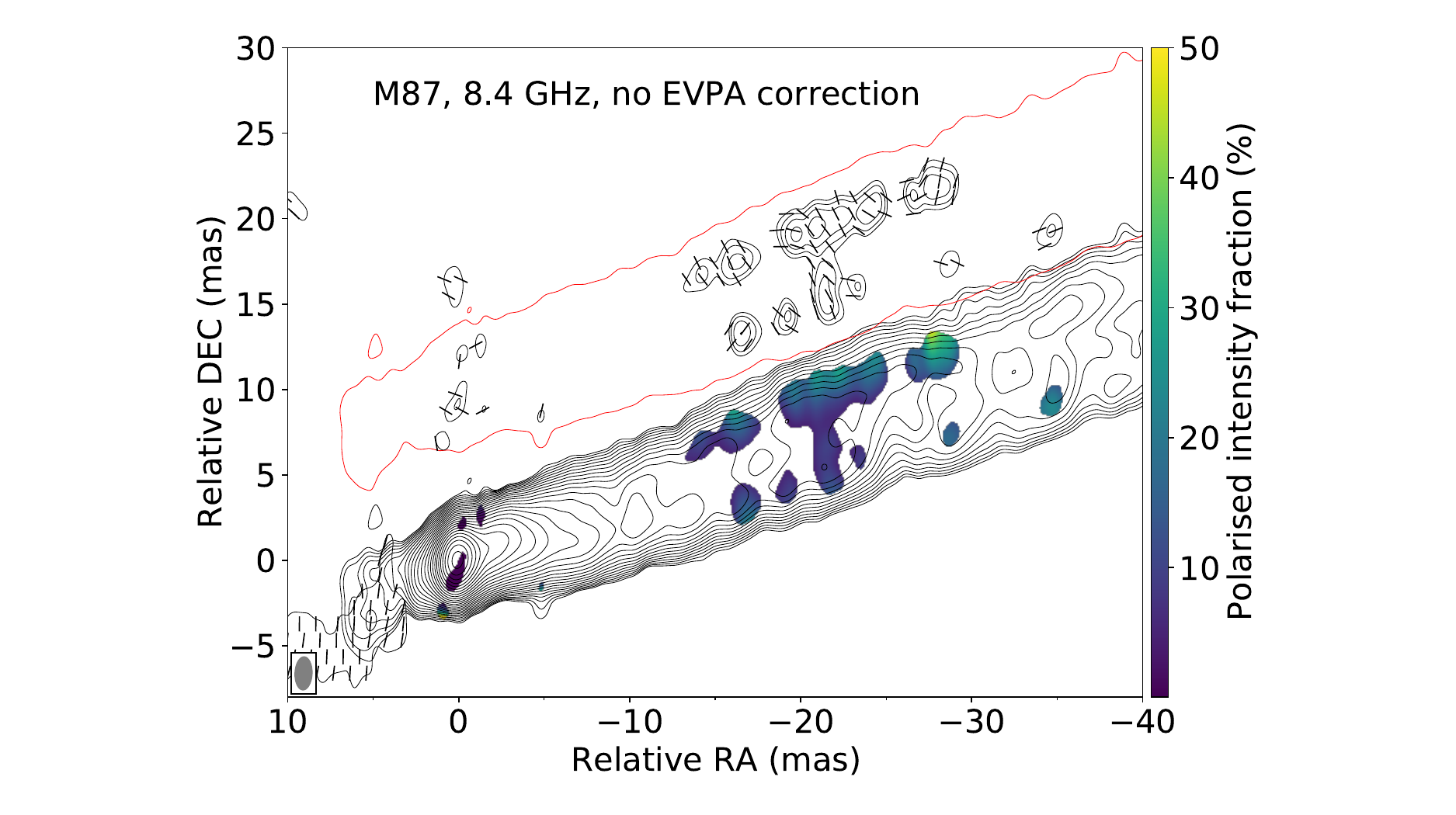}
	\includegraphics[width=0.48\linewidth, trim=5cm 0 4cm 0, clip]{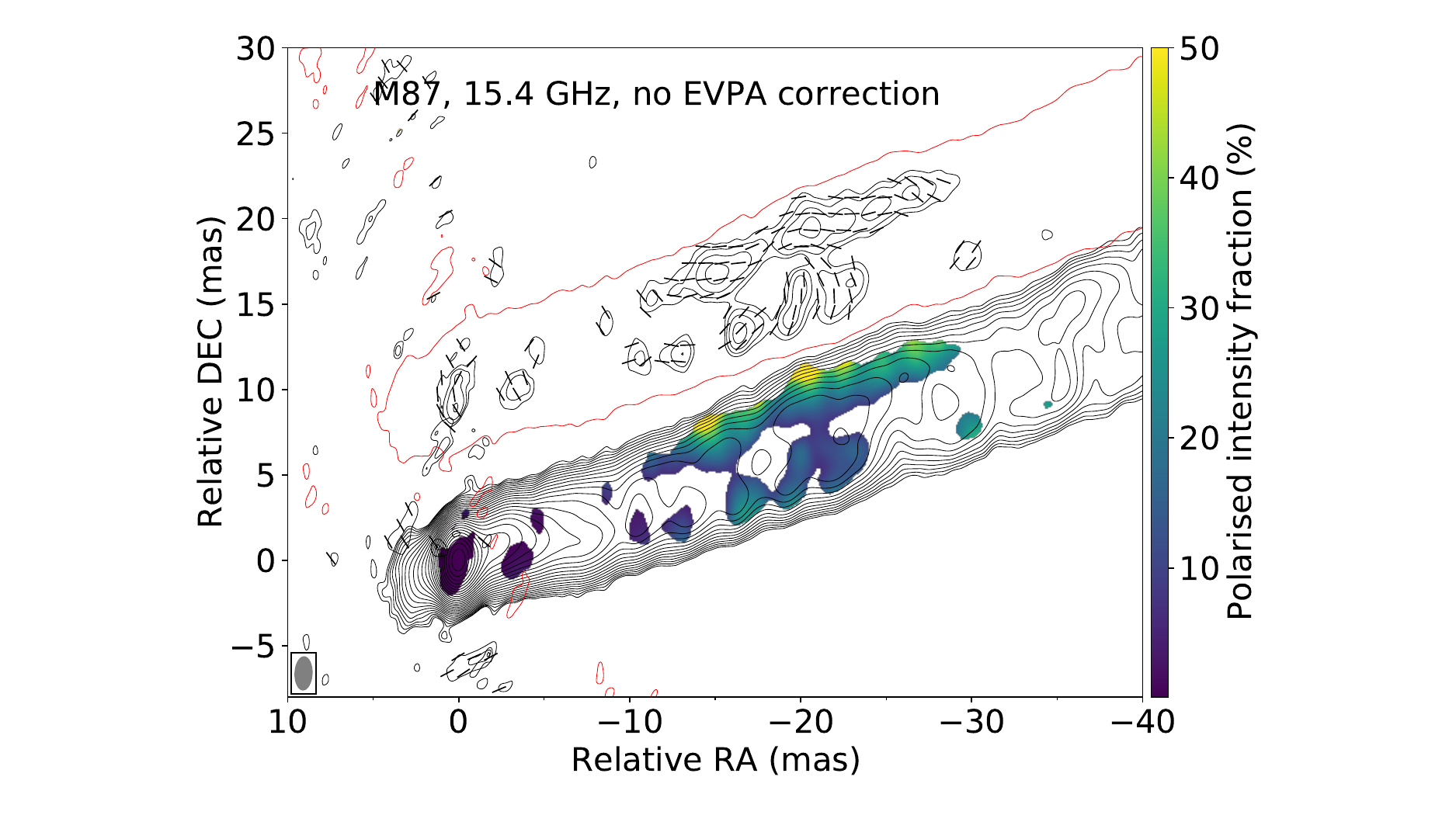}
	\includegraphics[width=\linewidth, trim=0.5cm 1.0cm 0.5cm 1cm, clip]{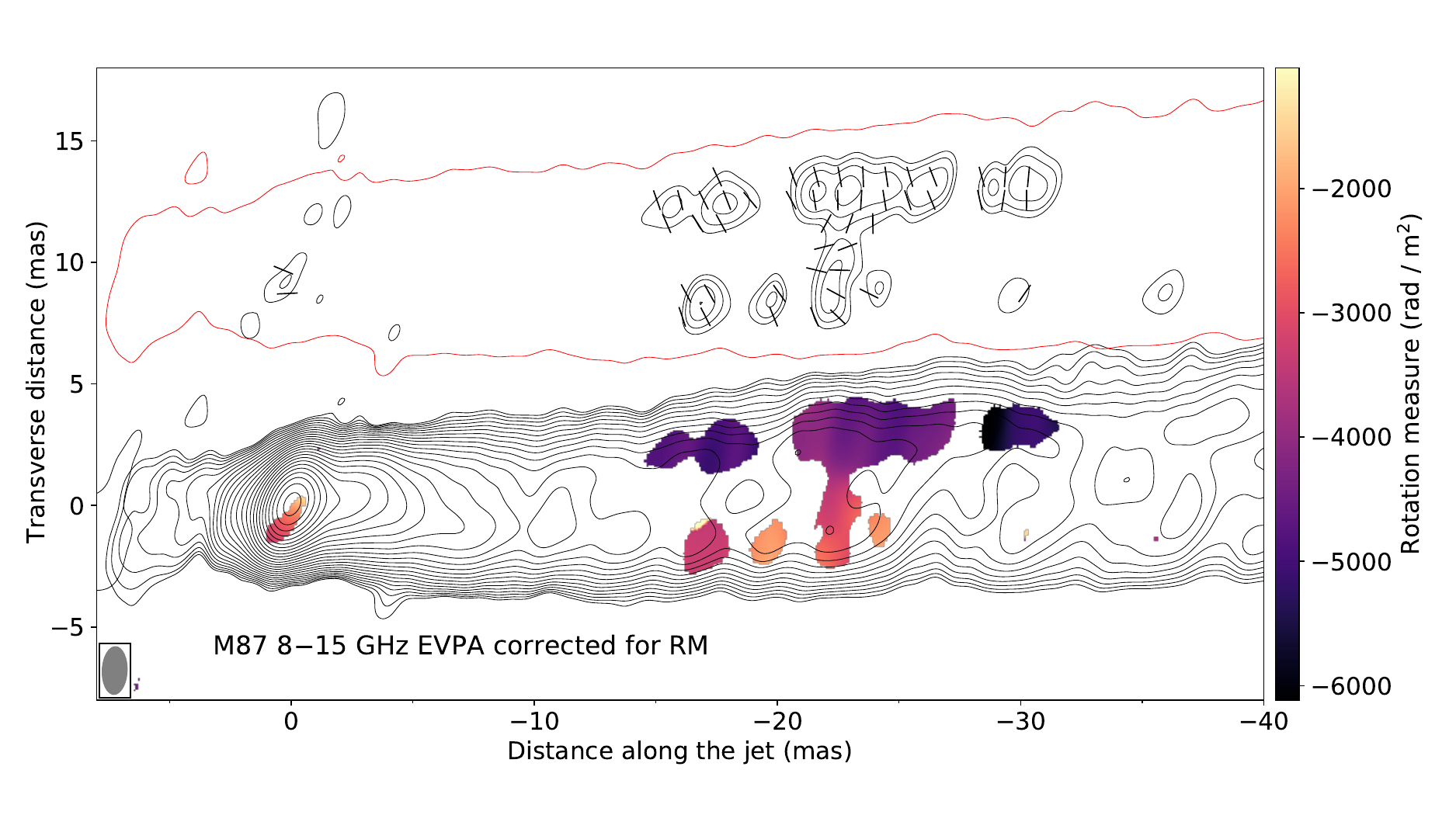}
    \caption{ 
    Polarisation images of M87 at 8\,GHz (upper left), at 15\,GHz (upper right) and Rotation Measure map between corresponding frequencies (bottom). The tick marks represent the polarisation position angle before (top) and after (bottom) correction for Rotation Measure. The black contours show the full intensity image levels which start from 260\,$\mu$Jy/beam at 8~GHz, 190\,$\mu$Jy/beam at 15\,GHz and increase by a factor of $\sqrt{2}$. All full intensity images convolved with $2\times1$\,mas, $\mathrm{PA}=-2^\circ$ elliptical beam with peak 1.22\,Jy/beam at 8\,GHz and 1.23\,Jy/beam at 15\,GHz. The single red contour represents the lowest full intensity level. The black contours inside the single red contour show polarisation intensity with peak 1.5\,mJy/beam at 8\,GHz and 1.8\,mJy/beam at 15\,GHz. The polarisation contours start at 440~$\mu$Jy/beam at 8\,GHz, 430\,$\mu$Jy/beam at 15\,GHz and increase by a factor of $\sqrt{2}$. The colours show polarised intensity fraction (top) and Rotation Measure (bottom).}
    \label{fig:stokes_qu}
\end{figure*}

\subsection{Jet to counter-jet flux ratio}
\label{sec:jet_to_counter-jet_flux_ratio}

The jet-to-counter-jet flux density ratio is a direct measurable value useful for estimating the basic physical parameters of relativistic jets such as viewing angle $\theta$ and Lorenz factor $\Gamma$. 
We measured the flux density of the counter-jet in the following way. The jet structure was modelled with several 2D Gaussian components, the brightest and closest to the phase centre was identified as the core. The core component only was convolved with the restoring beam. Then the obtained core image was subtracted from the original one, giving us well-separated jet and counter-jet structures. 
To get the flux ratio, we obtained fluxes from geometrically corresponding regions. For this, the length of the detected counter-jet was measured using the estimated in \autoref{sec:jet_shape} distance to the central SMBH. Thus, the boundary between the jet and the counter-jet was defined.

To measure viewing angle $\theta$ and Lorenz factor $\Gamma$, we used  
\begin{align}
\label{eq:jet_param}
\begin{split}
\theta &= \arccos \left( \frac{\xi - 1}{\sqrt{(4\beta_{\textrm{app}}^2+(\xi-1)^2 }} \right) , \\ 
\Gamma &= \frac{\xi+1}{\sqrt{2(\xi-\beta_{\textrm{app}}^2)}},  \\
\xi &\equiv \frac{\delta_\textrm{jet}}{\delta_\textrm{cjet}},
\end{split}
\end{align}
where  $  \delta_\textrm{jet}$, $\delta_\textrm{cjet} $ are the Doppler factors of the jet and the counter-jet, $\beta_\textrm{app}$ is the apparent jet speed \citep{Bottcher2011}. 
In the obtained images, continuous emission is observed, thus $\delta_\textrm{jet}/\delta_\textrm{cjet} = (   {F_{\textrm{jet}}}/{F_{\textrm{cjet}}} ) ^{\frac{1}{2-\alpha}} $, where ${F_{\textrm{jet}}}$ and $F_\textrm{cjet}$ are the fluxes of the jet and the counter-jet \citep{1979Natur.277..182S}. The estimations require a spectral index and apparent velocities. Since the length of observed counter-jet is roughly 4\,mas, the corresponding 4\,mas downstream the jet were considered. We take the spectral index of the jet from this region ($\alpha = -0.5 \pm 0.2$), which is the median value of the distribution in the map (\autoref{sec:sp_ind}). The jet speed was chosen as $ \beta_{\rm app} = 0.34 \pm 0.35$, which was estimated as an average of the apparent velocities of the jet features within 4 mas from the core \citep{2016AA...595A..54M}. Finally, we obtained results with different array configurations (VLBA+Y1; VLBA+Y1+Eb) and $\texttt{CLEAN}$ beams (elliptical; round). The averaged outcome is summarized here:
$$ 
\frac{F_{\rm jet}}{F_{\rm cjet}}=18\pm5,
\mbox{~}
\theta=17^\circ\pm8^\circ,
\mbox{~} 
\Gamma=1.2\pm0.1\,.
$$

\section{Linear polarisation and Rotation Measure}
\label{sec:RM}

Polarised emission moving through magnetized plasma can be affected by Faraday rotation. The effect can depolarise the light and change its EVPA. A relativistic jet is surrounded by a slow cocoon, optically active matter located in the magnetic field \citep{2021arXiv211104481S}, thus, in order to investigate the magnetic field in a jet, the effect should be taken into account \citep{1966MNRAS.133...67B, Hovatta_2012}.
The rotation measures $RM$ of extragalactic sources and can be used as a probe of the intergalactic medium to correct EVPA for Faraday rotation. The $RM$ has been determined by the linear fit:
\begin{equation}
    \phi_{\rm obs} = \phi_0 + RM \lambda^2,
\label{eq:phi}
\end{equation}
where $\phi_{\rm obs}$ is the observed polarisation angle, $\phi_0$ is the intrinsic EVPA and RM is the rotation measure. 
The $n \pi$ ambiguity of the $\phi_\mathrm{obs}$ measurements leads to uncertainties in $RM$ estimations. Thus,
\begin{equation}
    RM = \frac{\phi_{2} - \phi_{1} + n\pi}{\lambda_{2}^2 - \lambda_{1}^2}\,.
\label{eq:RM}
\end{equation}
To solve the $n \pi$ problem successfully, at least three frequencies are needed \citep{1981ApJS...45...97S}. Unfortunately, in our case, there are only two of them, so the problem of the $n \pi$ ambiguity is especially important.  
In \citet{Park2019}, it was shown that the rotation measure of the jet within 15 mas is changing in time though slowly, as compared to the measurement errors. So one can assume the stability of $RM$ over a long period of time, at least 20 years. Since the only unknown parameter in \autoref{eq:RM} is $n$, it can be selected to match $RM$ obtained using our data with the average value of $RM$, according to \citet{2002ApJ...566L...9Z} and \citet{Park2019}. The average $RM$ value for the period of 1995--2015 is $\approx -4500~\mathrm{rad/m}^2$. In addition, we calculated the $RM$ value, which provides full depolarisation with the condition of the $\pi$ EVPA rotation. This makes an upper bound for $|n|$. Finally, the average $RM$ and the maximum $|n|$ were used to solve the $n \pi$ ambiguity. For the case of $n=0$, the resulting mean $RM \approx -4000~\mathrm{rad/m}^2$ is consistent with previous measurements \citep{2002ApJ...566L...9Z, Park2019}. We assess the significance of selecting the integer value of $n$ by comparing the shift $\Delta RM$ associated with a change of $\Delta n = 1$ with the errors $\sigma_{\textrm{RM}}^{\textrm{old}} \approx 1000~\mathrm{rad/m}^2$ in the average $RM \approx -4500~\mathrm{rad/m}^2$ value reported in the literature, which we use as a reference point. The errors $\sigma_{\textrm{RM}}^{\textrm{old}}$ should be less than the shift $\Delta RM$, otherwise, multiple integer solutions for $n$ would exist, leading to ambiguity. In our particular case, a change of $\Delta n = 1$ corresponds to a shift of $\Delta RM \approx 2600$~$\mathrm{rad/m}^2$, which leads to a change of EVPA by $\Delta \phi \approx 60^{\circ}$. This supports the choice of $n$, since the errors of $RM$ measurement in \citet{2002ApJ...566L...9Z} and \citet{Park2019} are two times smaller. The galactic $RM$ correction was not applied to the M87 $RM$ map, since it is negligible for 8 and 15\,GHz \citep{2017MNRAS.467...83K}. We present the reconstructed deep M\,87 dual-frequency polarisation images and the Rotation Measure map in \autoref{fig:stokes_qu}. Assuming that the linear polarisation angle lies in the emitting particle orbit plane, i.e.\ perpendicular to the magnetic field lines, it is possible to reconstruct the magnetic field lines direction. In \autoref{fig:stokes_qu}, we observe that EVPA is oriented perpendicular to the jet direction near the edges, so the magnetic field lines are parallel to the jet axis. The opposite can be seen in the central jet regions. These results are consistent with the observations made by \citet{2002ApJ...566L...9Z} and \citet{Park2019}, where the perpendicular orientation of EVPAs in the northern edge of the M87 jet is observed. 

The Rotation Measure map (\autoref{fig:stokes_qu}) shows the difference between the northern and the southern limb of the jet. The value changes from approximately $-5000$ to $-2000$~rad/m$^2$. Having the mean value of $-4070 \pm 1030$~rad/m$^2$, the individual pixel errors are smaller $\sim$~300\,rad/m$^2$. The significance of the $RM$ gradient should be carried out carefully. Previous studies proposed several criteria for $RM$ gradient establishment \citep{2010ApJ...722L.183T, 2017MNRAS.467...83K}. Thus, the gradient of $RM$ in the map can be considered significant, since several criteria are satisfied:
\begin{enumerate}
    \item The M87 jet has a rich resolved transverse structure: more than three $\texttt{CLEAN}$ beams across the jet.
    \item A change in $RM$ is more than three times greater than the typical error.
    \item The gradient is located in an optically thin region.
    \item The $RM$ change is monotonic and smooth.
\end{enumerate}
Although the results show a smooth distribution of $RM$, it is important to note that the conclusion of the significance of the $RM$ gradient was made under the assumption of a uniform value of $n$ across the whole jet due to the limited frequency coverage. 

\section{Discussion} \label{sec:discussion}
\subsection{Jet morphology, spectral index distribution, and KH instability}

\begin{figure*}
    \centering
	\includegraphics[width=\linewidth,trim=0.5cm 7.2cm 0.5cm 0cm, clip]{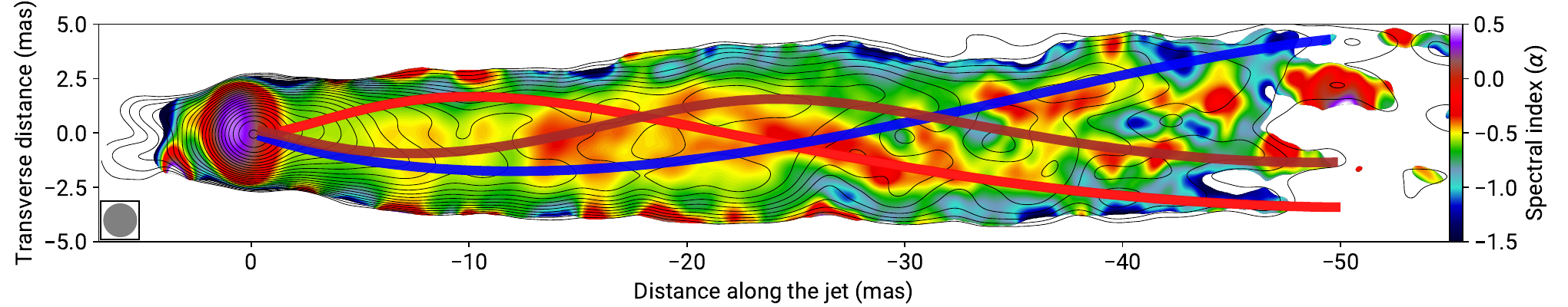}
    \caption{Threads identified from the Kelvin--Helmholtz instability modelling (red, blue and orange thick curves) overlaid on the bias-corrected spectral index map from \autoref{fig:sp_map_noEB_max} and the 8\,GHz intensity map, shown in contours which start from 450\,$\mu$Jy/beam increasing in a factor of $\sqrt{2}$ up to the peak value of 1.35\,Jy/beam. The 1.56-mas circular beam used for restoring both images is displayed in the bottom left corner at its FWHM level.  
    }
    \label{fig:sp_map_noEB_max_circ}
\end{figure*}

Morphological and spectral properties of the jet can be compared by overlaying the KH model threads from \autoref{sec:plasma_instability} onto the spectral index map from \autoref{fig:sp_map_noEB_max}. This overlay is shown in \autoref{fig:sp_map_noEB_max_circ} where one can see that the locations of the KH threads generally coincide with the regions of the flatter spectrum. This positional coincidence is in agreement with the theoretical expectation \citep{2000ApJ...533..176H,2006A&A...456..493P} for the elliptical surface mode and different body modes of KH instability to produce higher pressure regions near the jet boundary and in its interior. In these
regions, the combination of intrinsic heating \citep{2003NewAR..47..629L,2011ApJ...735...61H} and decreasing optical depth at higher frequencies should manifest itself with an apparent flattening of the synchrotron spectrum. 

Basic physical parameters of the jet and ambient medium can be derived from the mode identification described in \autoref{tab:KH_fit} by applying linear analysis of the KH instability \citep{2000ApJ...533..176H}. This allows us to estimate the jet Mach number, $M_\textrm{jet}$, the ratio, $\eta = h_\mathrm{jet}/h_\mathrm{ex}$ of specific enthalpy in the jet and the external medium, and the respective speeds of sound in the external medium and the jet, $a_{\textrm{ex}}$, $a_{\textrm{jet}}$. 
In these calculations, we use the estimates of the jet viewing angle, $\theta = 17{\fdg}2 \pm 3{\fdg}3$, apparent speed $v_\textrm{app} = (2.31 \pm 0.14)c$ and the pattern speed of instability $v_{w}/c = 0.34 \pm 0.21$ \citep{2016AA...595A..54M} and obtain M$_{\textrm{jet}}$ = 20$\pm$17, $\eta = 0.3 \pm 0.5$, $a_{\textrm{jet}} = 0.05 \pm 0.03$, $ a_{\textrm{ex}} = 0.03 \pm 0.01$. The resulting estimated Mach number and the enthalpy ratio are higher than what is typically expected in relativistic jets \citep{2008A&A...488..795R}, for example in a relativistic jet simulation for 3C~273 and 3C~31 show M$_{\textrm{jet}} \sim 3$, $\eta \sim 0.02$  \citep{2005MmSAI..76..110P, 10.1111/j.1365-2966.2007.12454.x}. Conversely, the respective sound speed in the jet plasma is lower than expected from those models. This apparent discrepancy may be explained by the effect of the dynamically important magnetic field which may affect the instability pattern \citep{2007ApJ...664...26H} but is not included in the plain KH models. Alternatively, it can result from underestimating the true jet speed from the apparent measured speed. At the viewing angle of the jet in M\,87, the detection of plasma motion at Lorentz factors $\gtrsim 3.38$ may be hindered by the differential Doppler boosting. Allowing for this effect, we can adopt the highest detected speed in M87 \citep[$v_\textrm{app} \approx 6c$;][]{1999ApJ...520..621B} for our estimates, which results in M$_{\textrm{jet}} \approx 5$, $\eta \approx 0.014$, and $a_{\textrm{jet}} \approx 0.24$. These values are in good agreement with the estimates obtained for the kiloparsec-scale jet in M87 \citep{2003NewAR..47..629L}.

The edge brightening of the jet which can be seen in \autoref{fig:stokes_i_no_EB} and also has been reported previously \citep{2016ApJ...833...56A, Walker, 2018A&A...616A.188K, 2019A&A...626A..75J} can be explained by the stratification of a jet in velocity, density, and internal energy \citep{Walker, 2021A&A...654A..27B} or by the presence of dynamically important magnetic field \citep{2021NatAs...5.1017J, 2021A&A...656A.143K}. 
Each of these effects should also provide a flattening of the spectrum toward the geometrical axis of the jet. In the first case, an interaction between layers in a velocity-stratified jet can heat the plasma, increasing the internal energy and the spectral index in the spine region. In the second case, the spectral index can increase due to opacity variations induced by the magnetic field \citep{2011MNRAS.415.2081C}. A combined action of these factors is also plausible \citep{2018A&A...616A.188K}, but our present results do not warrant a more detailed assessment of the physical conditions causing the observed edge brightening of the jet.

The observed spectral index distribution presented in \autoref{fig:sp_map_noEB_max} indicates that the emission in the compact region at the base of the jet is optically thick at frequencies below 15\,GHz, most likely due to synchrotron self-absorption. The estimated errors of the spectral index are high in this region, making it difficult to locate the transition from optically thick to thin emission. With these errors taken into account, the transition may be occurring in the jet anywhere between $\approx -1.1$\,mas and $\approx +1.5$\,mas axial separation from the coordinate origin.
On the counter-jet side, the spectral index is reliably measured at separations exceeding $\approx 1.5$\,mas, allowing for comparison of the emission properties in the jet and the counter jet. The median values of the spectral index distribution for the jet and the counter-jet regions are $\alpha_\mathrm{j} = -0.5 \pm 0.2$, $ \alpha_\mathrm{cj} = -0.8 \pm 0.2$. The spectral index can also be estimated for the HST-1 feature, using the measured integral flux densities $F_{\textrm{8GHz}} \approx 19$ mJy and $F_{\textrm{15GHz}} \approx 8$~mJy, 
which yields $\alpha_\mathrm{HST-1} \approx -1.4$. Although this estimate should formally indicate pronounced synchrotron ageing of the emission, it should be viewed with extra caution because, at the location of HST-1, the 15\,GHz flux density may be underestimated which would cause spurious steepening of the measured spectral index.

\subsection{Magnetic field structure}

Similarly to the total intensity maps, the images of fractional linear polarisation in \autoref{fig:stokes_qu} are also edge brightened. For the optically thin emission, this effect can result from apparent or true depolarisation in the central regions of the jet caused by a helical magnetic field \citep{2021Galax...9...58G} or a turbulent plasma flow \citep{2014ApJ...780...87M}.
The possibility for such a structure to result from a $\texttt{CLEAN}$ imaging artefact due to the residual uncleaned polarized flux \citep{10.1093/mnras/stad525} is not likely, as we employed a deep $\texttt{CLEAN}$ in this work \autoref{sec:sp_ind}.

The variations of the EVPA shown in \autoref{fig:stokes_qu} can be reconciled with in a hollow, edge brightened jet \citep{2023MNRAS.523..887F} seen at a viewing angle $\sim 1/\Gamma$ and threaded by a large scale helical magnetic field with the pitch angle $>45^\circ$ in the plasma frame \citep{2005MNRAS.360..869L}. Similar EVPA morphology can be obtained for the force-free reverse field pinch \citep{2011MNRAS.415.2081C} and pure helical magnetic field model, both analytically \citep{2013MNRAS.430.1504M,10.1093/mnras/stad121} and numerically \citep{2021A&A...656A.143K}.

Some indications in favour of the presence of a substantial regular magnetic field in the jet may also be found in the transverse gradient of the Faraday rotation measure observed in the jet. 
Although transverse gradients of the rotation measure may as well be produced by the differences in the density of a Faraday screen of thermal electrons, which surrounds the jet, the polarization images of the large jet in M\,87 \citep{Pasetto_2021} also suggest the presence of a helical magnetic field.
The observed anti-correlation between the side of the jet with higher degrees of polarization and the side with higher RM magnitude \autoref{fig:stokes_qu} could also point at the helical magnetic field \citep{2021Galax...9...58G}.

\section{Summary}\label{sec:summary}

This paper presents an investigation of the physical properties of the parsec-scale jet in M\,87 obtained from imaging the jet with augmented VLBA at 8 and 15 GHz. The main observational results of this work are:
\begin{enumerate}
    \item Images of total and polarised intensity are obtained at each frequency. The total intensity images, reaching a record dynamic range >20000:1 show edge-brightening, faint counter-jet, HST-1 knot and reveal helical threads in the jet. 
    \item $\texttt{CLEAN}$ bias, which strongly affects a spectral index map obtained from the individual total intensity images, is identified and corrected for. The bias-corrected spectral index map demonstrates a complex pattern.
    \item The linear polarisation maps uncover the change of magnetic field lines from edges to the jet's centre.
    \item The rotation measure map shows a significant gradient perpendicular to the jet direction.
\end{enumerate}

A detailed analysis of the observational information obtained from the VLBA data yields the following conclusions: 
\begin{enumerate}
    \item The helical threads observed in the jet can be explained by the Kelvin--Helmholtz instability in the jet. This interpretation is also supported by the spectral index map, where the flattening of the spectra traces well the observed helical threads.
    \item The edge brightening observed in the total intensity and fractional polarisation can be interpreted either by transverse velocity stratification of relativistic plasma or by a large-scale helical magnetic field.  
    \item The faint structure southeast to the core is confirmed as the counter-jet. According to the spectral index map analysis, this feature is optically thin $\alpha_\mathrm{cj} = -0.8 \pm 0.2$ and hence this region cannot be the jet origin. 
    \item Intrinsic physical jet parameters are estimated from modelling the observed jet structure: the jet viewing angle $\theta = 17^{\circ} \pm 8^{\circ}$, Lorenz-factor $\Gamma = 1.2 \pm 0.1$, expansion index $k = 0.532 \pm 0.008$, Mach number $M_\textrm{jet}$ = 20$\pm$17, jet to ambient medium density ratio $\eta = 0.3 \pm 0.5$ and the deprojected distance from the VLBI core to the SMBH $r_\textrm{c} = (5 \pm 2) \times 10^{-2}$\,pc.
\end{enumerate}

Our results reveal that Kelvin-Helmholtz instability starts to develop in the regions relatively close to the central engine of the jet ($\sim10^{2}\textrm{--}10^{4}R_{g}$). This specific region corresponds to the jet's formation and collimation zone \citep[e.g.,][]{2018ApJ...868..146N,2020MNRAS.495.3576K}, emphasizing the significant role of plasma instability in the jet morphology and evolution. Consequently, it is crucial to consider plasma instability when studying jet properties. The next step for the investigation will be the study of the temporal evolution of the helical pattern in the jet. However, due to the limited sensitivity of modern telescopes, it is challenging to track the helical pattern in survey mode. Thus, only rare full-track observations can be utilized for the analysis. The situation for M87-type jet studies will be greatly improved by ngVLA \citep{2018ASPC..517...15S,2018ASPC..517....3M}. In addition to the sensitivity, short baselines will provide an opportunity to close the gap between parsec and kilo-parsec scale jets.

\section*{Acknowledgements}

We thank an anonymous referee for careful reading and valuable feedback, that have greatly enhanced the clarity of this paper. We thank M.~L.~Lister, D.~C.~Homan, K.~I.~Kellermann, E.~E.~Nokhrina, M.~Janssen, J. Livingston, J.~I.~Nikonova and J. R\"oder for useful discussions and comments as well as Elena Bazanova for language editing.
This research was supported in part by the Russian Science Foundation (project 16-12-10481).
This work is part of the M2FINDERS project which has received funding from the European Research Council (ERC) under the European Union’s Horizon 2020 Research and Innovation Programme (grant agreement No 101018682).
A.~S.~Nikonov received financial support for this research from the International Max Planck Research School (IMPRS) for Astronomy and Astrophysics at the Universities of Bonn and Cologne.
The National Radio Astronomy Observatory is a facility of the National Science Foundation operated under cooperative agreement by Associated Universities, Inc. 
This work is partly based on the observations with the 100-m telescope of the MPIfR (Max-Planck-Institut f\"ur Radioastronomie) at Effelsberg.
This research has made use of the data from the University of Michigan Radio Astronomy Observatory which has been supported by the University of Michigan and by a series of grants from the National Science Foundation, most recently AST-0607523. 

\smallskip
Facilities: VLBA, VLA, Effelsberg telescope.

Software: Astropy \citep{astropy:2013, astropy:2018, astropy:2022}, Matplotlib \citep{Hunter:2007}, Numpy \citep{harris2020array}, Python \citep{10.5555/1593511}, Scipy \citep{2020SciPy-NMeth}.

\section*{Data availability}

The correlated data underlying this article are accessible from the public NRAO archive. The 8 and 15~GHz Stokes I and polarisation intensity images, the original and the corrected for bias spectral index maps are available from the article and as FITS files in its online supplementary material. The maps are also available from the public MOJAVE archive\footnote{\url{https://www.cv.nrao.edu/MOJAVE}}. 



\bibliographystyle{mnras}
\bibliography{lib}


\bsp	
\label{lastpage}
\end{document}